\documentclass[11pt]{article}

\usepackage[a4paper,margin=1in]{geometry}

\usepackage{pb-diagram}
\usepackage{latexsym}
\usepackage[usenames,dvipsnames]{xcolor}
\usepackage{amsmath,amsfonts,amssymb,amsmath,epsfig}
\usepackage[english]{babel}
\usepackage{graphicx}
\usepackage[font=small,labelfont=bf]{caption}
\usepackage{bm}
\usepackage{tikz-cd}
\usepackage[hidelinks,linktoc=all]{hyperref}
\usepackage{cite}
\usepackage{physics}
\numberwithin{equation}{section}

\usepackage{tocloft}
\makeatletter
\g@addto@macro\bfseries{\boldmath}
\makeatother

\makeatletter
\def\blfootnote{\gdef\@thefnmark{}\@footnotetext}
\makeatother

\usepackage{pb-diagram}
\usepackage[usenames,dvipsnames]{xcolor}
\usepackage{amsmath,amsfonts,amssymb,amsmath,epsfig}
\usepackage{graphicx}
\usepackage{bm}
\usepackage{tikz-cd}

\newcommand\mailto[1]{\href{mailto:#1}{\tt #1}}

\def\cD{{\cal D}}

\newcommand{\be}{\begin{equation}}
\newcommand{\ee}{\end{equation}}

\def\nonu{\nonumber}
\def\br{\begin{eqnarray}}
\def\er{\end{eqnarray}}

\def\({\left(}
\def\){\right)}
\def\[{\left[}
\def\]{\right]}

\def\lie{{\cal G}}
\def\a{\alpha}

\def\g{\gamma}

\def\k{\kappa}
\def\l{\lambda}

\def\o{\omega}

\def\pa{\partial}

\def\t{t}

\def\tp0{\Thet\mathcal{A}_{+}^{(0)}}
\def\tm0{\Thet\mathcal{A}_{-}^{(0)}}

\def\l{\lambda}

\def\nonu{\nonumber}

\def\bi{\begin{itemize}}
\def\ei{\end{itemize}}

\newcommand{\kdv}{\text{KdV}}
\newcommand{\mkdv}{\text{mKdV}}
\newcommand{\mkdvI}{\text{mKdV-I}}
\newcommand{\mkdvII}{\text{mKdV-II}}
\newcommand{\vac}{\text{vac}}
\newcommand{\dI}{\partial_x^{-1}}

\DeclareMathAlphabet{\mathpzc}{OT1}{pzc}{m}{it}

\renewcommand{\u}{\mathcal{J}}
\renewcommand{\v}{\mathpzc{V}}

\newcommand{\uu}{\mathcal{J}_1}
\newcommand{\vv}{\mathpzc{V}_1}
\newcommand{\phii}{\phi_1}
\newcommand{\etaa}{\eta_1}
\newcommand{\ub}{\mathcal{J}_2}
\renewcommand{\vb}{\mathpzc{V}_2}
\newcommand{\phib}{\phi_2}
\newcommand{\etab}{\eta_2}

\newcommand{\bb}{\mathfrak{B}}

\begin{document}

\renewcommand{\thefootnote}{\fnsymbol{footnote}}
\pagenumbering{gobble}
\thispagestyle{empty}

\vspace*{1.5cm}

\begin{center}

{\LARGE\bfseries
Negative flows of  generalized KdV and mKdV \\[.6em] hierarchies
and their gauge-Miura transformations
}\\[1.0cm]

{\bfseries
Ysla F. Adans,${}^{\!a}$\footnote{\mailto{ysla.franca@unesp.br}}~%
Guilherme  Fran\c ca,${}^{\!b}$\footnote{\mailto{guifranca@gmail.com}}~%
Jos\' {e} F. Gomes,${}^{\!a}$\footnote{\mailto{francisco.gomes@unesp.br}} \\[.5em]
Gabriel V. Lobo,${}^{\!a}$\footnote{\mailto{gabriel.lobo@unesp.br}}~and~%
Abraham H. Zimerman${}^{a}$\footnote{\mailto{a.zimerman@unesp.br}}
}\\[0.5cm]

${}^{a}${\itshape Institute of Theoretical Physics --- IFT/UNESP}\\
{\itshape Rua Dr. Bento Teobaldo Ferraz 271, 01140-070, S\~ ao Paulo, SP, Brazil}\\[.5em]
${}^{b}${\itshape University of California, Berkeley, CA 94720, USA}\\[1.0cm]

{\bf Abstract}
\end{center}
\vspace{-1em}

The KdV hierarchy is a paradigmatic example of the rich mathematical structure underlying  
integrable systems and has  far-reaching connections in several areas of  theoretical physics. 
While the positive part   of the KdV hierarchy  is well known, 
in this paper we consider an affine Lie algebraic construction for its negative part.
We show that the original Miura transformation can be extended
to a gauge transformation that implies several new types of relations among the negative flows
of the KdV and mKdV hierarchies. Contrary to the positive flows, 
such a ``gauge-Miura'' correspondence becomes degenerate whereby 
more than one  negative mKdV model is mapped into a single negative KdV model.
For instance, the sine-Gordon  and another negative mKdV flow 
are mapped into a single negative KdV flow which  inherits solutions of both former models.
The gauge-Miura correspondence implies a rich degeneracy regarding solutions
of these hierarchies. 
We  obtain similar results for the  generalized KdV and mKdV hierachies constructed
with the affine Lie algebra $\widehat{s\ell}(r+1)$.
In this case the first negative mKdV flow corresponds to an affine Toda field theory  
and the gauge-Miura correspondence yields its KdV counterpart. In particular, 
 we show explicitly a KdV analog of the Tzitz\'{e}ica-Bullough-Dodd model.
In short, we uncover a rich mathematical structure for the negative flows of integrable hierarchies obtaining novel relations and integrable systems.

\vfill

\newpage
\tableofcontents
\newpage

\renewcommand*{\thefootnote}{\arabic{footnote}}
\setcounter{footnote}{0}
\pagenumbering{arabic}
\setcounter{page}{1}

\section{Introduction}

The KdV equation is perhaps the first example of an integrable model 
and its  study has led to several remarkable
relations in mathematical  physics.
Indeed, the modern theory of inverse scattering transform
was originally developed for the  KdV equation  
\cite{Miura_1968,Gardner_1967,Gardner_1974},
and later extended to the nonlinear Schr\" odinger equation
\cite{Zakharov_1972} as well as to several other  important integrable models.
A striking connection between these techniques  
and the Bethe ansatz allowed the development of the quantum inverse
scattering transform \cite{Sklyanin:1978,Faddeev_1979,Sklyanin_1982},
with important applications in statistical mechanics of lattice systems and nonperturbative methods in quantum  field theory \cite{Korepin_book}.

It became clear that the KdV equation fits into a much more general structure, namely an \emph{integrable
hierarchy} of nonlinear differential equations  \cite{Lax_1968,Gelfand_1975,Miwa_2000}.
In fact, more general integrable hierarchies can be constructed 
from a zero curvature condition and systematically classified in terms of Kac-Moody algebras  
\cite{Drinfeld_1985,Leznov_1983,Olive_1985,Olive_1993,Olive_1993b,Babelon_1991,Babelon_1993,Babelon_2003,Aratyn_1991,Groot_1992,Hollowod_1993,Miramontes_1999,Ferreira_1997}.
The KdV hierarchy, and related models suchs as mKdV, sine-Gordon, Liouville theory, affine Toda field theories, 
etc., appear 
in numerous areas of theoretical physics
such as 2D CFTs
\cite{Zamolodchikov_1985,Sasaki_1988,Eguchi_1989,Bazhanov_1996} and
string theory
\cite{Polyakov_1981,Douglas_1990,Gross_1990,Banks_1990,Witten_1991,Kontsevich_1992,Dijkgraaf_1991,Dijkgraaf_1992,Itzykson_1992}.
It is also worth noting that there is a deep --- and not fully understood --- connection between classical and quantum
integrability that goes beyond a  classical
limit \cite{Krichever_1997,Bazhanov_2014}, e.g.,  the
generating function of quantum transfer matrices in spin chains
can be identified with tau-functions of classical integrable models
\cite{Alexandrov_2013}, besides their equivalence to the partition function
of a matrix model formulation of 2D  quantum and topological gravity  \cite{Witten_1991,Kontsevich_1992,Dijkgraaf_1992,Dijkgraaf_1991}.
Recently, the KdV hierarchy has been  
the central object  in a number of papers, such as 
thermal correlation functions of 2D   CFTs 
\cite{Maloney_2019,Maloney_2019b,Dymarsky_2019,Dymarsky_2019b},
eigenstate thermalization hypothesis \cite{Dymarsky_2019c,Dymarsky_2019d},
and black holes on 
 $\text{AdS}_3$ 
\cite{Perez_2016,Erices_2019,Dymarsky_2020,
Grumiller_2020,Cardenas_2021,Lenzi_2021,Lenzi_2023}.

A cornerstone of the inverse scattering transform
is the \emph{Miura transformation}  that links the 
KdV and mKdV equations besides
establishing  a map to  a Schr\" odinger spectral problem~\cite{Miura_1968}.
Since both equations are just one member of their respective hierarchies,
a natural question concerns the relation between the other models --- or flows --- 
of  these hierarchies.
Indeed, the algebraic construction of the KdV and mKdV hierarchies for general
(untwisted) affine Lie algebras, together with a generalization of the Miura
transformation, is well-established \cite{Groot_1992}.
It
has also recently been shown that the Miura transformation can be seen
as  a gauge transformation that maps all \emph{positive flows} of the KdV and
mKdV hierarchies into each other
\cite{Ana_2016,Lobo_2021}.
However, these integrable hierarchies also admit \emph{negative flows}, 
which often turn out
to be  (nonlocal) integro-differential equations. The first negative
flow of some integrable hierarchies are of particular interest since
they correspond to a relativistic affine Toda field theory \cite{Aratyn_1991}, such as the sine-Gordon model.

While the negative odd \cite{Miramontes_1999} and negative even \cite{Franca_2009} algebraic structures of the mKdV hierarchy are known, 
the negative part of the KdV hierarchy has not been  previously considered.
It is the goal of this paper to provide the affine Lie algebraic construction
of the negative part of the KdV hierarchy, 
and  moreover to show that a \emph{gauge-Miura} transformation
provides a map among the negative flows.
However, this relation becomes degenerate, namely more than
one negative mKdV flow maps into a single negative KdV flow.
Such a correspondence also leads to interesting identities
 besides the typical Miura transformation, which we call ``temporal Miura transformations;'' they are not present  for the positive part
 of these hierarchies.
Ultimately, such a mapping can be traced back to the action of dressing operators
on two types of vacuum, zero and nonzero (constant), defining two separate sets
of mutually commuting flows of the mKdV hierarchy.

The standard KdV and mKdV hierarchies are constructed in terms of the affine  Lie algebra $\widehat{s\ell}(2)$, which is a particular case of a more
general construction  \cite{Groot_1992}.
Thus, we also extend explicitly the aforementioned connections to $\widehat{s\ell}(3)$, yielding new integrable models
such as  KdV counterparts of  the affine two-component Toda field theory and 
Tzitz\' eica-Bullough-Dodd model.
We then generalize these results more abstractly 
to $\widehat{s\ell}(r+1)$.
We find that the gauge-Miura transformations increase the degeneracy for the negative flows, i.e.,
$r+1$ mKdV models are mapped into a single KdV model.
We also point out an interesting relation between the vacuum structure of 
these models, and how a single
generalized mKdV-type of solution generates several 
generalized KdV-type of solutions.

This paper is organized as follows.
In sec.~\ref{sec:pos_flow} we introduce the positive flows of the KdV and mKdV hierarchies
under  similar construction in terms of $\widehat{s\ell}(2)$. 
This provides a unifying perspective between them and facilitates
their gauge-Miura correspondence.  
In sec.~\ref{sec:mkdv_neg} we introduce the negative mKdV flows, the most notable example
being the sine-Gordon model, and show that negative even flows only admit solutions with a nonzero vacuum that implies a deformation  on the dressing construction of solitons 
via deformed vertex operators.
In sec.~\ref{sec:kdv_neg} we propose an algebraic construction for the negative part of the KdV hierarchy. 
In sec.~\ref{sec:gauge_neg_flows} we show how the gauge-Miura transformations lead to degenerate
relations between these hierarchies, which can be classified according to a zero or nonzero vacuum configuration.  Indeed, in sec.~\ref{sec:heis} we show that such vacua lead to two
distinct sets of mKdV commuting flows. 
These connections are  further generalized explicitly 
to $\widehat{s\ell}(3)$ 
in sec.~\ref{sec:extension} and  abstractly to $\widehat{s\ell}(r+1)$
in sec.~\ref{sec:extension2}.
Along the way, new integrable models as well as new  
relations among existing models arise, such as
a KdV counterpart of the relativistic Tzitz\' eica-Bullough-Dodd model. 
Background material on affine (Kac-Moody) algebras are summarized in the
 appendix~\ref{sec:algebra}.

\section{Positive KdV and mKdV flows}
\label{sec:pos_flow}

In this section we  introduce  the KdV and mKdV hierarchies under a similar 
affine Lie algebraic construction. Although these hierarchies can be introduced in different ways, e.g., by  the AKNS construction or in terms of the Lax equation with pseudodifferential operators, 
our construction
allows us to establish interesting connections between them systematically.
We  also emphasize how the Miura transformation can
be extended to a gauge transformation between the positive flows 
of these hierarchies
\cite{Ana_2016,Lobo_2021}. Latter on this connection will  be generalized to the negative flows.
(For details on affine Lie algebras we refer to the appendix~\ref{sec:algebra}.)

Consider the affine Lie algebra 
$\lie = \widehat{s\ell}(2)$ under the principal gradation.
The mKdV spatial gauge potential is defined as 
\be
\label{Ax_mkdv}
A_x^{\mkdv} \equiv E + A_{0} = 
\begin{pmatrix} \v & 1\\  \l & - \v \end{pmatrix}  ,
\ee
where $A_{0} \equiv  \v(x,t) h^{(0)} \in \lie_{0}$  contains
the field $\v$ and $E \equiv E_{\a}^{(0)} + E_{-\a}^{(1)} \in \lie_{1}$ is a semisimple element.  
Similarly, the KdV hierarchy can be constructed under this very same algebraic structure
but with the gauge potential 
\be 
\label{Ax_kdv}
A_x^{\kdv} \equiv E + A_{-1} = \begin{pmatrix} 0 & 1 \\ \l + \u & 0 \end{pmatrix}  ,
\ee
where now the field $\u$ is associated to 
$A_{-1} \equiv  \u(x,t) E_{-\a}^{(0)} \in \lie_{-1}$.  
An important connection between the two  hierarchies is the gauge-Miura transformation
\be
\label{gauge}
\begin{split}
A_x^{\kdv} &= S  A_x^{\mkdv} S^{-1} + S \pa_x S^{-1} .
\end{split}
\ee
There exist two operators satisfying this equation \cite{Lobo_2021}, namely
$S$ can be either $S_1$ or $S_2$  given by
\be
S_{1} = \begin{pmatrix}
  1 & 0\\ 
  \v & 1
\end{pmatrix} , \qquad 
S_{2} = \begin{pmatrix}
  0 & \l^{-1}\\ 
  1 & -\l^{-1}\v
\end{pmatrix} 
=  
\begin{pmatrix}
  1 & 0 \\ 
  -\v & 1 
\end{pmatrix} \lambda^{-1} E
.
\label{miura_gauge}
\ee
They yield the following  relations between the KdV and mKdV fields:
\be \label{miura}
\u =  \v^2  \mp \pa_x \v .
\ee
The minus sign comes from $S_1$ and the plus sign from $S_2$.
Eq.~\eqref{miura}  is the seminal Miura transformation \cite{Miura_1968},
originally introduced as a map
between solutions of the mKdV equation into solutions of the KdV equation. 
This  transformation 
played a fundamental role in the development of the inverse scattering transform \cite{Gardner_1967,Gardner_1974} and it
is also important  
on a quantum level, e.g., in connection to CFTs \cite{Bazhanov_1996}.
The gauge transformation \eqref{gauge}  lifts the Miura transformation 
to a mapping between the entire positive parts of the mKdV and KdV hierarchies (this will be made 
explicit shortly).

Let $A_x$ and $A_{t_N}$ denote a pair of  gauge potentials.
Integrable hierarchies can be  constructed  from the \emph{zero curvature condition}  \cite{Aratyn_1991,Groot_1992,Hollowod_1993,Ferreira_1997,Miramontes_1999}
\be
[ \pa_x + A_x, \pa_{t_N} + A_{t_N} ] = 
\pa_xA_{t_N}  - \pa_{t_N}A_{x }  + [ A_x, A_{t_N} ] = 0 ,
\label {zero_curvature}
\ee
where  $N$ indexes a ``time flow,'' i.e., each $N$ gives rise to one
nonlinear integrable model described by a partial differential equation.  The algebraic structure of the hierarchy
is uniquely specified by $A_x$ while $A_{t_N}$ must be a sum of suitable graded operators.
For instance,
for the mKdV hierarchy defined by \eqref{Ax_mkdv} we have
\be 
A_{t_N}^{\mkdv}  = D^{(N)}_N + D^{(N-1)}_N + \cdots+ D^{(0)}_N 
\qquad \big(D^{(n)}_N \in \lie_n \big), \label{At_mkdv}
\ee
whereas for the KdV hierarchy \eqref{Ax_kdv} 
we  have
\be 
A_{t_N}^{\kdv}  = { \cal D}^{(N)}_N+ {\cal D}^{(N-1)}_N + \cdots + {\cal D}^{(0)}_N + {\cal D}^{(-1)}_N \qquad \big({\cal D}^{(n)}_N \in \lie_n\big). 
\label{At_kdv}
\ee
Importantly, the  zero curvature equation~\eqref{zero_curvature} decomposes 
as a consequence of the grade structure of the  algebra, 
specified by a suitable grading operator,
allowing us to solve for each  $D^{(n)}_N$ and  $\mathcal{D}^{(n)}_N$ nontrivially. 
In particular, the the highest grade  component 
yields
\be \label{highest_mkdv}
\big[ E, D^{(N)}_N \big]  = 0
\ee
and
\be 
\big[ E, {\cal D}^{(N)}_N \big]  = 0  ,
\label{highest_kdv}
\ee
besides $\pa_x D^{(N)}_N = 0$ and $\pa_x \mathcal{D}^{(N)}_N = 0$,
respectively,
implying that both $D^{(N)}_N$ and ${\cal D}^{(N)}_N$  are constant elements lying
in the kernel of $E$, 
denoted by $\mathcal{K}_E \equiv \big\{ X \, | \, [X,E]  = 0\big\}$ --- see eq.~\eqref{Kernel} and
note that $\mathcal{K}_E$ has only odd graded elements.
It therefore follows  that $N$ must be \emph{odd}, 
i.e., $N=2n+1$
for $n=1,2,\dotsc$. This is the reason why the \emph{positive parts} of both  KdV and mKdV hierarchies only
admit equations of motion associated to \emph{odd time flows}.  
The lower grade components then solve for  
each remaining $D^{(n)}_N$ and $\mathcal{D}^{(n)}_N$ recursively.
For  mKdV, the zero grade component finally  yields the
Leznov-Saveliev equation \cite{Leznov_1983}
\be 
\pa_{t_N} A_0 -\pa_xD^{(0)}_N -\big[ A_0, D^{(0)}_N \big]=0 ,
\label{zero_grade_mkdv}
\ee
which is the equation of motion for  the field  $\v(x,t_N)$
parametrizing  $A_0$.
On the other hand, for the KdV hierarchy the  
equation of motion is  obtained from the $-1$ grade component
\be
\pa_{t_N} A_{-1} -\pa_x{\cal D}^{(-1)}_N - \big[ A_{-1}, {\cal D}^{(0)}_N \big]=0 ,
\label{minus_one_grade_kdv}
\ee
since  the field $\u(x, t_N)$ is  associated to  $A_{-1}$. 
In this manner all the nonlinear differential equations within these hierarchies
are systematically obtained from the algebraic structure
of  the spatial gauge potential $A_x$.

Concretely, with the differential operator 
$\mathcal{P} \equiv \pm \partial_x - 2 \v$, 
the first positive  flows of the KdV  and mKdV hierarchies, as well as their
equivalence under  gauge-Miura  
\eqref{gauge}--\eqref{miura}, are 
described as follows.

\begin{itemize}
\item \underline{$t_{1}$-flow} 
\be
(\pa_{t_1} - \pa_x) \u  = \mathcal{P}(\pa_t - \pa_x) \v =0 .
\ee
This case yield chiral wave equations on both sides, showing
that $t_1 = x$.

\item \underline{$t_{3}$-flow} 
\be
4 \pa_{t_3} \u - \pa_x^3 \u + 6 \u \pa_x\u = \mathcal{P}
 \left( 4 \pa_{t_3} \v - \pa_x^3 \v + 6 \v^2 \pa_x \v \right)
= 0 .
\label{kdv_mkdv_t3}
\ee
On the LHS  we recognize the celebrated  KdV equation, while on the RHS   we recognize the mKdV equation, which name their respective hierarchies.

\item \underline{$t_{5}$-flow} 
\begin{multline}
16 \pa_{t_5}\u - \pa_x^5\u + 20 \pa_x \u \pa_x^2 \u + 10 \u \pa_x^3 \u - 30 \u^2 \pa_x \u  = 
\mathcal{P} \big( 16 \pa_{t_5} \v  - \pa_x^5 \v \\ + 40 \v \pa_x \v \pa_x^2 \v + 10 \v^2 \pa_x^3 \v
+ 10 (\pa_x\v)^3 - 30 \v^4 \pa_x\v \big) = 0 .
\label{kdv_mkdv_t5}
\end{multline}
 On the LHS  we recognize the 
 Swada-Kotera equation \cite{Sawada_1974,Caudrey_1976}
 and on the RHS we have its modified counterpart in the mKdV hierarchy
 \cite{Fordy_1980}.

\item \underline{$t_{7}$-flow} 
\be
\begin{split}
&64 \pa_{t_7} \u - \pa_x^7\u + 70 \pa_x^2\u \pa_x^3 \u + 42 \pa_x\u \pa_x^4 \u - 70 (\pa_x\u)^3 
\\
& \qquad 
+ 14 \u \pa_x^5 \u -280 \u \pa_x \u \pa_x^2 \u   - 70 \u^2 \pa_x^3 \u    + 140 \u^3 \pa_x \u  
\\  &=  
\mathcal{P}
\Big(
64 \pa_{t_7} \v - \pa_x^7\v + 14  \v^2 \pa_x^5\v  + 84 \v \pa_x\v \pa_x^4 \v  + 140 \v \pa_x^2 \v \pa_x^3 \v   
\\ 
& \qquad\qquad + 126 (\pa_x \v)^2 \pa_x^3\v    - 70 \v^4 \pa_x^3 \v  + 182 \pa_x \v (\pa_x^2 \v)^2  
\\ 
& \qquad \qquad - 560 \v^3 \pa_x \v \pa_x^2 \v - 420 \v^2 \pa_x^3\v  + 140 \v^6 \pa_x\v
\Big) = 0.
\end{split}
\ee

\item The above pattern repeats itself for every  higher-order partial differential equation
within these hierarchies ($N=9, 11, \dotsc$).

\end{itemize}

The gauge-Miura transformation \eqref{gauge} thus
provides a 1-to-1 correspondence between the positive flows of the mKdV and
KdV hierarchies as summarized by the diagram
\be \label{corresp_pos}
\begin{tikzcd}[row sep=0.2em, column sep=2em]
t_{N}^{\mkdv} \arrow[r, "S"] & t_{N}^{\kdv} 
\end{tikzcd}
\ee
for $N = 2n+1$,  $n=0,1,\dotsc$, and where $S$ can be any of the two choices given in eq.~\eqref{miura_gauge}.
Note that  since there are two gauge transformations, leading to two 
different Miura transformations~\eqref{miura}, a \emph{single} mKdV-solution
generates \emph{two} possible KdV-solutions between associated models.

\section{Negative mKdV flows}
\label{sec:mkdv_neg}

The negative  flows of the mKdV hierarchy
have been previously considered \cite{Franca_2009}.
In this case  
the temporal gauge potential has the form  
\be 
A_{t_{-N}}^{\mkdv}  = D^{(-N)}_{-N}+ D^{(-N+1)}_{-N}+\cdots D^{(-1)}_{-N} \qquad \big(D^{(-n)}_{-N} \in \lie_{-n}\big)
\label{At_mkdv_neg}
\ee
and   leads to  a series of --- usually nonlocal --- equations of motion that are systematically obtained from the zero curvature condition
\be
\big[ \pa_x + A_{x}^{\mkdv} ,  \pa_{t_{-N}} + A_{t_{-N}}^{\mkdv} \big] =0.
\label{mkdv_zcc_neg}
\ee
Again, this equation
 decomposes into graded components that can be solved recursively, but  now  
starting from the lowest grade
\be \label{mkdv_lowest_neg}
\pa_x D^{(-N)}_{-N} + \big[ A_0, D^{(-N)}_{-N} \big]=0 
\ee
which fixes  $D^{(-N)}_{-N}$. The second lowest component yields $D^{(-N+1)}_{-N}$, and so on, until
the zero grade component yields  the equation of motion
\be
\pa_{t_{-N}}  A_0 - \big[ E, D^{(-1)}_{-N} \big] =0 .
\label{mkdv_neg_eq}
\ee
It is important to note that, contrary to the positive part of the mKdV hierarchy,  
the solution  to eq.~\eqref{mkdv_lowest_neg} no longer requires  $D^{(-N)}_{-N}$ to lie in the kernel 
$\mathcal{K}_E$. Thus,   
no constraint is enforced on the admissible values of $N$, i.e., 
the \emph{negative flows}
of the mKdV hierarchy can be both \emph{odd and even}.
We provide a few examples below.
\begin{itemize}
\item  \underline{$t_{-1}$-flow} 
\be 
 \pa_{t_{-1}} \pa_x \phi =  
 2\sinh ( 2 \phi ). 
 \label{sg_eq}
\ee
This is the well-known sinh-Gordon model in  light cone coordinates.\footnote{On can identify $t_{-1} = \bar{z}$ and $x = z$ as the light cone coordinates.
Under the transformation 
$t = \tfrac{ z+\bar{z}}{a \sqrt{2}}$,
$x = \tfrac{ z-\bar{z}}{ a \sqrt{2}}$, 
and 
$\phi = \tfrac{i \beta}{2} \varphi$, with $a^2 = \pm \tfrac{\alpha_0}{16}$,
one obtains from \eqref{sg_eq} the sine-Gordon model with  Lagrangian
$\mathcal{L} = \tfrac{1}{2} \partial_\mu \varphi \partial^\mu \varphi  \mp \tfrac{\alpha_0}{\beta^2}\big( \cos(\beta \varphi) - 1\big)$.
}
In solving  the zero curvature equation one finds
\be
A_{t_{-1}}^{\mkdv}= e^{-2 \dI \v} E_{\a}^{(-1)} +e^{2 \dI \v } E_{-\a}^{(0)},    
\qquad 
\pa_{t_{-1}} \v =  e^{2\dI  \v} - e^{-2\dI \v} .
\label{sg_potential}
\ee
We then define the operator%
\footnote{The definition \eqref{dIDef}  
is  the inverse derivative
operator obeying $\pa_x \pa_x^{-1} f = \pa_x \int f dx = f$ and $\pa_x^{-1} \pa_x f = \int \pa_x f dx = f$ for any function $f$.
}
\be \label{dIDef}
\dI f \equiv \int f  dx 
\ee
such that $\pa_x \pa_x^{-1} = \pa_x^{-1} \pa_x = 1$,
and to obtain \eqref{sg_eq} from \eqref{sg_potential} we introduce 
a simple change of the field variable%
\footnote{
This relation comes from the group
parametrization $A_0 = B^{-1} \pa_x B$, which in this
case is $B = e^{\phi h^{(0)}}$.}
\be \label{vdphi}
\v \equiv \pa_x \phi .
\ee

\item \underline{$t_{-2}$-flow}
\be  \label{mkdv_minus_two_eq}
\dfrac{1}{2} \pa_{t_{-2}}   \pa_x \phi = - e^{-2 \phi} \dI e^{2 \phi} - e^{2 \phi} \dI e^{-2 \phi} .
\ee
The temporal gauge potential in this case is
\be
A_{t_{-2}}^{\mkdv} = h^{(-1)}+ 2  e^{-2 \phi}\dI e^{2 \phi} E_{\a}^{(-1)}  
- 2 e^{2 \phi} \dI e^{-2 \phi}  E_{-\a}^{(0)} .
\label{mkdv_two_potential} 
\ee 

\item \underline{$t_{-3}$-flow} 
\be \label{mkdv_minus_three_eq}
\dfrac{1}{2} \partial_{t{-3}} \partial_x \phi = 
e^{2\phi} \dI \[ e^{-2\phi} \dI\big( e^{2\phi} - e^{-2\phi} \big)  \] 
+ e^{-2\phi} \dI \[ e^{2\phi} \dI \big( e^{2\phi} - e^{-2\phi} \big)  \] .
\ee

\item \underline{$t_{-4}$-flow} 
\be \label{mkdv_minus_four_eq}
\begin{split}
\dfrac{1}{4} \pa_{t_{-4}}\pa_x \phi &= 
e^{-2\phi} \dI\left[ e^{2\phi} \dI\big( e^{-2\phi} \dI e^{2\phi} + e^{2\phi} \dI e^{-2\phi}  \big)  \right] \\
& - 
e^{2\phi} \dI\left[ e^{-2\phi} \dI\big( e^{-2\phi} \dI e^{2\phi} + e^{2\phi} \dI e^{-2\phi}  \big)  \right] .
\end{split}
\ee

\item One can proceed for $-N = -5, -6, \dotsc$ 
to obtain  higher-order integro-differential equations within
the negative part of the mKdV hierarchy.\footnote{%
Some of these equations may be written in local form by further
differentiation, e.g., equation~\eqref{mkdv_minus_two_eq} can be written as
$ \pa_{t_{-2}} \pa_x^2 \v - 4 \v^2 \pa_{t_{-2}}\v - (\v)^{-1} (\pa_x\v)  (\pa_{t_{-2}} \pa_x \v) 
- 4 (\v)^{-1} \pa_x \v = 0$.}

\end{itemize}

At this point let us mention
a peculiar feature of the negative part of the mKdV hierarchy concerning the \emph{vacuum} \cite{Franca_2009}.
The equations of motion associated to odd and even flows have 
qualitatively different type of solutions. 
Solitons  are constructed in the orbit of 
some  vacuum \cite{Ferreira_1997}, thus 
different vacua  generate  
 different types of solutions.
For the \emph{positive flows} of
the mKdV hierarchy --- see  eqs.~\eqref{kdv_mkdv_t3} and
\eqref{kdv_mkdv_t5} ---
the zero vacuum $\v_0 = 0$ is clearly a solution, and so is
 a constant vacuum $\v_0 \ne 0$.  
However,  for the \emph{negative flows} the situation is different.
Indeed,  $\v_0 = 0$  is   a solution
of both the sinh-Gordon  \eqref{sg_potential} 
and  eq.~\eqref{mkdv_minus_three_eq}, but a constant vacuum
$\v_0 \ne 0$  is not a solution of these models.
On the other hand,  the zero vacuum $\v_0 = 0$ is neither  a solution of  
eq.~\eqref{mkdv_minus_two_eq} nor eq.~\eqref{mkdv_minus_four_eq}, although a constant
vacuum   $\v_0 \ne 0$ is a solution to both models. More specifically,  the factor
\be
e^{2 \pa_x^{-1} \v_0 } - e^{-2 \pa_x^{-1} \v_0} = 
e^{2 \v_0 x } - e^{-2 \v_0 x} = 0
\ee
only for $\v_0 = 0$. This term  appears in  all \emph{negative odd} equations. On the other hand, the factor
\be
e^{-2 \dI \v_0} \dI e^{2 \dI \v_0} + e^{2 \dI \v_0} \dI e^{-2 \dI \v_0 } 
=
e^{-2 \v_0 x} \dfrac{e^{2 \v_0 x}}{2\v_0}
+ e^{2\v_0 x} \dfrac{ e^{-2\v_0 x} }{(-2\v_0)}
= 0 
\ee
only for $\v_0 \neq 0$. This term appears in all \emph{negative even} equations.
In fact,  all  models associated to \emph{negative even  flows only admit nonzero vacuum solutions}, while all models associated to
\emph{negative odd flows  only admit zero vacuum solutions}.
This can be seen by considering
the zero curvature equation at the vacuum configuration:
\be
\left[ A_{x, \text{vac}}^{\mkdv},   A_{t_{-N}, \text{vac}}^{\mkdv} \right]   =    
\left[ E + \v_0 h^{(0)} ,   D^{(-N)}_{-N,\text{vac}}+D^{(-N+1)}_{-N, \text{vac}}+\cdots+D^{(-1)}_{-N,\text{vac}} \right]  = 0. 
\label{vacc}
\end{equation}
If $\v_0 \neq 0$ the lowest grade equation is 
$\big[ \v_0 h^{(0)} ,   D^{(-N)}_{-N,\text{vac}}\big]  = 0$,
implying that  $D^{(-N)}_{-N,\text{vac}}$ commutes  with $h^{(0)}$ and therefore  from 
eq.~\eqref{graded_subspaces2} we see that $D^{(-N)} _{-N,\text{vac}} \in \lie_{-2n}$, i.e.,  $N=2n$. 
However   if $\v_0 = 0$ then the lowest  grade projection becomes
$\big[ E ,   D^{(-N)}_{-N,\text{vac}} \big]  = 0$,
implying that $ D^{(-N)}_{-N,\text{vac}} \in \mathcal{K}_E \subset \lie_{-2n+1}$ which only admits
odd-graded elements, 
i.e.,   
 $N=2n-1$.\footnote{These restrictions do not apply to the positive part of the mKdV hierarchy because the highest grade component is always
$[E^{(1)}, D^{(N)}_{N,\text{vac}}] = 0$ (recall eq.~\eqref{highest_mkdv}), implying that $N$ is odd and both zero
and nonzero vacua are allowed.}
In short:
\begin{itemize}
\item The positive part of the mKdV hierarchy has only odd 
flows and its integrable models  
admit solutions related to both zero ($\v_0 = 0$) and nonzero ($\v_0 = \mbox{const.} \ne 0$) vacuum.
\item The negative part of the mKdV hierarchy  splits into two subhierarchies, one indexed by even 
 flows whose models  only admit strictly nonzero vacuum  ($\v_0 = \mbox{const.} \ne 0$), and the other indexed by 
odd flows whose models  only
 admit zero vacuum ($\v_0 = 0$).  
\end{itemize}
In sec.~\ref{sec:heis} we will revisit and explain in more detail the role of the vacuum,  showing how
they generate two separate sets of commuting flows that define
an integrable hierarchy.

\section{Negative KdV flows}
\label{sec:kdv_neg}

For the negative part of the KdV hierarchy 
we have the Lax operator \eqref{Ax_kdv}  and we now
propose
\be
A_{t_{-N}}^{\kdv} = \cD^{(-N-2)}_{-N} + \cD^{(-N-1)}_{-N} + \cdots + \cD^{(-1)}_{-N} \qquad \big( \cD^{(-n)}_{-N} \in \lie_{-n}\big) . 
\label{At_kdv_neg}
\ee
The zero curvature condition \eqref{zero_curvature} decomposes according to the grade structure of
the algebra --- in the principal gradation --- yielding
\begin{subequations}\label{neg_decomp}
\begin{align}
\big[ A_{-1}, \cD^{(-N-2)}_{-N} \big] &= 0 , \label{lowestkdv} \\
\pa_x \cD^{(-N-2)}_{-N} + \big[A_{-1}, \cD^{(-N-1)}_{-N} \big] &= 0 , \\
&\ \, \vdots \nonumber \\
\pa_x \cD^{(-1)}_{-N} + \big[E, \cD^{(-2)}_{-N} \big] - \pa_{\t_{-N}} A_{-1} &= 0 , \label{eqofmotionkdv} \\
\big[ E, \cD^{(-1)}_{-N} \big] &= 0. \label{ker}
\end{align}
\end{subequations}
We can solve for each $\cD^{(-n)}_{-N}$ recursively and the  equation of motion with respect to the time evolution parameter   $\t_{-N}$ is given by \eqref{eqofmotionkdv}.
Note that since $A_{-1} = \u(x, t_{-N}) E_{-\a}^{(0)}$  the lowest grade eq.~\eqref{lowestkdv} 
  implies that   $\cD^{(-N-2)}$ is proportional to $E_{-\a}^{(-m)}$, therefore $N = 2m-1$.
Thus, \emph{the KdV hierarchy only admits negative odd flows}.
This is in contrast  to the  mKdV case previously discussed where $N$ can take both even and  odd 
  negative  values. This will play an important role later on when we discuss  gauge
 transformations between the negative part of these hierarchies.

Similarly to the mKdV case, 
the equations of motion for the negative part of the KdV hierarchy are  
more conveniently expressed in terms of the
field $\eta$ defined by
\be
\u \equiv \pa_x \eta. \label {eta}
\ee

\begin{itemize}
\item \underline{$t_{-1}$-flow} 
\be 
\pa_{t_{-1}} \pa_x^3  \eta - 4 \pa_x\eta \pa_{t_{-1}} \pa_x  \eta
- 2  \pa_x^2 \eta \, \pa_{t_{-1}} \eta = 0 .
\label{kdv_neg1}
\ee
This equation is the counterpart of the sinh-Gordon model but in the KdV hierarchy. It  is
obtained by 
solving eqs.~\eqref{neg_decomp} with $N=1$,  yielding the temporal gauge potential
\be
A_{\t_{-1}}^{\kdv} = \frac{\pa_{t_{-1}}\eta}{2}  \left( E_{\a}^{(-1)} + E_{-\a}^{(0)}  \right)
    + \frac{\pa_{t_{-1}}\pa_x\eta}{4}  \; h^{(-1)}
    + \frac{ 2\pa_{t_{-1}} \eta \, \pa_x\eta - \pa_{t_{-1}}\pa_x^2\eta }{ 4 } \; E_{-\a}^{(-1)} .
 \label{At_kdv_minus1}
\ee
As we will show, solutions of the sinh-Gordon  \eqref{sg_eq} and
also of model \eqref{mkdv_minus_two_eq} generate  solutions
to model \eqref{kdv_neg1} via Miura transformations. 
Recall that there are  two possible gauge-Miura transformations so this model inherits four possible solutions from the mKdV hierarchy.

\item \underline{$t_{-3}$-flow} 
\be
\begin{split}
& -(1/2) \pa_{t_{-3}} \pa_x^5 \eta - 4 \pa_x \eta \left(  2 \pa_{t_{-3}} \pa_x \eta \, \pa_x \eta  
 - \pa_{t_{-3}} \pa_x^3  \eta  + \pa_x^2\eta \, \pa_{t_{-3}} \eta \right) \\ 
& + 5 \pa_x^2\eta \, \pa_{t_{-3}}\pa_x^2  \eta  
+4 \pa_{t_{-3}} \pa_x  \eta \, \pa_x^3\eta + \pa_x^4\eta \, \pa_{t_{-3}}\eta 
\\
& - \pa_x^2 \eta \, 
\dI \left(4 \pa_x\eta \, \pa_{t_{-3}} \pa_x \eta + 2 \pa_x^2 \eta  \, \pa_{t_{-3}} \eta 
- \pa_{t_{-3}}\pa_x^3 \eta \right)=0.
\end{split}
\label{kdv_neg3}
\ee
This equation is obtained by  solving eqs.~\eqref{neg_decomp} with $N=3$ yielding 
\be
\label{At_kdv_minus3}
\begin{split}
A_{\t_{-3}}^{\kdv} &= \frac{\pa_{t_{-3}}\eta}{2} \left( E_{\a}^{(-1)} + E_{-\a}^{(0)} \right) 
    + \frac{ \pa_{t_{-3}} \pa_x \eta}{4}  h^{(-1)}
    - \frac{{\cal B}}{8}  \left( E_{\a}^{(-2)} + E_{-\a}^{(-1)} \right) \\
    &+ \frac{ 2 \pa_{t_{-3}}\eta \pa_x\eta - \pa_{t_{-3}}\pa_x^2\eta }{ 8 } \; E_{-\a}^{(-1)} - \frac{{\pa_x\cal B}}{16} \; h^{(-2)} + \frac{ \pa_x^2{\cal B} - (\pa_x\eta)  {\cal B} }{ 8 } \; E_{-\a}^{(-2)}, 
\end{split}
\ee
where
\be
{\cal B} \equiv \dI  \left(4 \pa_x \eta \pa_{t_{-3}} \pa_x  \eta +2 \pa_x^2 \eta \pa_{t_{-3}} \eta -\pa_{t_{-3}}\pa_x^3 \eta \right) .
\ee

\item One can proceed systematically in this fashion to obtain  lower negative KdV flows,
but the equations quickly become  complicated.

\end{itemize}

A few remarks are warranted.
The nonlinear model~\eqref{kdv_neg1}  first appeared in  \cite{Verosky_1991} and was obtained through Olver's inverse recursion operator.%
\footnote{Originally, this equation was written as $\pa_t \u = \pa_x w$ and $\pa_x^3 w + 4 \u \pa_x w + 2 (\pa_x \u )w = 0$,
which is equivalent to~\eqref{kdv_neg1} with $\u = - \pa_x \eta$.}
This model is known to be related to the Camassa-Holm equation by a reciprocal transformation 
\cite{Fuchssteiner_1996} and a  more natural equivalence with the
associated Camassa-Holm equation has also been noted \cite{Hone_1999}.
Several properties of this model have already been  
studied  \cite{Qiao_2012}, such as its bi-Hamiltonian structure, conservation laws, Hirota bilinear
transformation,  soliton and quasi-periodic solutions.
The above derivation provides the affine algebraic construction from
which this model arises.

By a similar argument as that used with  eq.~\eqref{vacc} to analyze the
possible vacuum solutions we now conclude:\footnote{%
For the positive flows of the KdV hierarchy the zero curvature 
condition at the vacuum yields
$\big[E, \mathcal{D}_{N,\vac}^{(N)}\big] = 0$ as the highest grade component,
regardless whether $\u_0 = 0$ or $\u_0 \ne 0$;
in both cases $\mathcal{D}_{N,\vac}^{(N)}$ is in the kernel of $E$ and
thus $N$ is odd.
For the negative flows, 
$\big[A_{x,\vac}^{\kdv}, A_{t_{-N,\vac}}^{\kdv}\big] = 
\big[ E + \u_0 E_{-\alpha}^{(0)}, 
\mathcal{D}_{-N,\vac}^{(-N-2)} + \dotsm \mathcal{D}_{-N,\vac}^{(-1)} \big] = 0$,
whose lowest grade component is 
$\big[E, \mathcal{D}_{-N,\vac}^{(-N-2)}\big] = 0$ for $\u_0 = 0$, and
$\big[\u_0 E_{-\alpha}^{(0)}, \mathcal{D}_{-N,\vac}^{(-N-2)}\big] = 0$ 
for $\u_0 \ne 0$.
In the former case  $D_{-N,\vac}^{(-N-2)}$ is in the kernel 
of $E$ which only  admits $-N$ odd, whereas in the latter
case $D_{-N,\vac}^{(-N-2)} \sim E_{-\alpha}^{(m)}$ which also implies
that $-N$ is odd. Thus, for the KdV hierarchy, zero and nonzero vacua
are admissible for all flows, namely  positive odd and negative odd. 
}
\begin{itemize}
\item Each integrable model within the negative part of the KdV hierarchy admits both zero (\mbox{$\u_0 = 0$}) as
    well as nonzero ($\u_0 = \mbox{const.} \ne 0$) vacuum solutions. 
\end{itemize}
This  behavior differs  from the negative mKdV hierarchy which splits 
into negative odd and negative even flows,  
separately admitting zero or nonzero vacuum solutions, respectively.

\section{Gauge transformation for negative flows}
\label{sec:gauge_neg_flows}

In sec.~\ref{sec:pos_flow} we saw that the entire positive part of  the KdV and mKdV hierarchies
are related by --- the same --- gauge transformation; see diagram \eqref{corresp_pos}.
 A critical question is how to extend this correspondence to the negative
part of these hierarchies. Recall that   mKdV splits into negative 
even and negative odd  flows, while  KdV has only negative odd flows. Therefore, there is a mismatch in the number of equations to begin with and the correspondence seems a priori ambiguous.  
Next, we  show that this apparent contradiction is in fact resolved by  careful consideration of the gauge-Miura transformation, and an interesting structure emerges.

Let us start with the transformations
\be
S_1(\phi)   =  e^{(\pa_x\phi)E_{-\a}^{(0)}}  =   \begin{pmatrix}
  1 & 0\\ 
  \pa_x\phi & 1
\end{pmatrix},
\qquad
S_2 = S_1(-\phi) E^{(-1)} ,
\label{g}
\ee
which we know already connects the spatial gauge potentials  $ A_{x}^{\mkdv}$  and  $A_{x}^{\kdv}$, namely
\be \label{gauge2}
A_{x}^{\kdv}= S A_{x}^{\mkdv} S^{-1} + S\partial_{x} S^{-1}  
= E_{\a}^{(0)}+E_{-\a}^{(1)}+ \u E_{-\a}^{(0)} .
\ee
This holds true for either  $S = S_1$ or $S = S_2$. Each choice realizes one respective Miura transformation ($\v = \pa_x\phi$ and $\u = \pa_x\eta$):
\begin{equation}
\pa_x \eta  = (\pa_x\phi)^2 \mp \pa_x^2\phi .
\label{miura2}
\end{equation}
Importantly in our following argument is that,
under such  gauge transformations
a \emph{pair} of temporal mKdV gauge potentials, $A_{t_{-2n+1}}^{\mkdv}$  and $A_{t_{-2n}}^{\mkdv} $, coalesce into a \emph{single}
temporal KdV gauge potential, $A_{t_{-2n + 1}}^{\kdv}$.

Let us first consider  $ A_{t_{-2n+1}}^{\mkdv}$ under
the gauge transformation induced by $S_1$.\footnote{Analogous results can be obtained with $S_2$ and
are presented in the appendix~\ref{sec:alternative_gauge}.}
Since  the operator $D^{(-2n+1)}_{-2n+1} \sim E_{\a}^{(-n)}$ the  gauge operator $S_1$ in eq.~\eqref{g}  yields
 \be
 \begin{split}
 A_{\t_{-2n+1}}^{\kdv} &=  S_1 \(  D^{(-2n+1)}_{-2n+1}+ D^{(-2n+2)}_{-2n+1}+\cdots D^{(-1)}_{-2n+1}\) S_1^{-1} + S_1 \pa_{t_{-2n+1}}S_1^{-1}  \\
 &= {\cal D}^{(-2n-1)}_{-2n+1} + {\cal D}^{(-2n)}_{-2n+1} +  {\cal D}^{(-2n+1)}_{-2n+1} +\cdots + {\cal D}^{-1}_{-2n+1} \label{kdv1} .
\end{split}
\ee
 Similarly, since   $D^{(-2n)}_{-2n} \sim h^{(-n)}$  the transformation for $ A_{t_{-2n}}^{\mkdv}$ yields
 \be
\begin{split}
 \widetilde{A}_{\t_{-2n+1}}^{\kdv} &\equiv  S_1 \(  D^{(-2n)}_{-2n}+ D^{(-2n+1)}_{-2n}+\cdots D^{(-1)}_{-2n}\) S_1^{-1} - S_1 \pa_{t_{-2n}}S_1^{-1} \\
 &= {\widetilde {\cal D}}^{(-2n-1)}_{-2n+1} + {\widetilde {\cal D}}^{(-2n)}_{-2n+1} +  {\widetilde {\cal D}}^{(-2n+1)}_{-2n+1} +\cdots + {\widetilde {\cal D}}^{-1}_{-2n+1} .
 \end{split}
 \label{kdv2}
\ee
Now, the potentials $A_x^{\kdv}$ and $ A_x^{\mkdv} $ are universal within the hierarchies. Therefore  
the zero curvature  condition for \eqref{kdv1} and \eqref{kdv2} must yield the same
operator because they have the same graded algebraic structure. In other words, the zero curvature
condition together with $A_x$ uniquely fixes all integrable models within the hierarchy. Thus,  
\be \label{equal_op}
{\cal D}^{(-j)}_{-2n+1} = {\widetilde {\cal D}}^{(-j)}_{-2n+1},  \qquad  {A}_{\t_{-2n+1}}^{\kdv}=\widetilde{A}_{\t_{-2n+1}}^{\kdv}  ,
\ee
and both gauge potentials must provide the  same  evolution equations.
 We therefore conclude that \emph{subsequent negative odd and even mKdV flows collapse into the same negative odd KdV flow}. This is depicted by
 the diagram
\be\label{corresp_neg}
\begin{tikzcd}[row sep=0.1em, column sep=3em]
t_{-N}^{\mkdv} \arrow[dr, "S"] & \\
&  t_{-N}^{\kdv} \\ 
t_{-N-1}^{\mkdv} \arrow[ur, swap, "S"] 
\end{tikzcd}
\ee
for $N = 2n - 1$, $n=1,2,\dotsc$, and where $S$ can be either
$S_1$ or $S_2$ from eq.~\eqref{g}.
Such a 2-to-1 correspondence should be compared with the 1-to-1 correspondence
\eqref{corresp_pos} for the positive part
of these hierarchies.
The above  relation also explains why each negative KdV flow admits both zero and also nonzero vacuum
solutions:
\begin{itemize}
\item A zero (nonzero) vacuum solution of a given negative KdV flow is inherited from
a solution of the associated negative odd (even) mKdV flow.
Interestingly, \emph{two different mKdV models} yield different types of solution
to the \emph{same KdV model}.
Moreover, we have two possible Miura transformations,
each yielding a different solution of the KdV model.
\end{itemize}

Let us consider explicitly the first negative KdV flow, which according to 
diagram \eqref{corresp_neg} is related to the first two negative mKdV flows.  The gauge
transformation \eqref{kdv1} yields  
\be
\label{tm11}
\begin{split}
 A_{\t_{-1}}^{\kdv} &= S_1 A_{t_{-1}}^{\mkdv} S_1^{-1} +S_1\partial_{t_{-1}} S_1^{-1} \\
    &= e^{ (\pa_x \phi) E_{-\a}^{(0)}}\left(e^{-2\phi}E_{\a}^{(-1)}+e^{2\phi}E_{-\a}^{(0)}\right) e^{-(\pa_x\phi) E_{-\a}^{(0)}}  - \pa_x \pa_{t_{-1}}\phi E_{-\a}^{(0)} \\
&= e^{-2\phi}\left(E_{\a}^{(-1)}+E_{-\a}^{(0)}\right) - \pa_x \phi \, e^{-2\phi}  h^{(-1)}-(\pa_x \phi)^2 e^{-2\phi}E_{-\a}^{(-1)} .
\end{split}
\ee
Comparing \eqref{tm11} with \eqref{At_kdv_minus1} we conclude that the identity 
\be \label{mt1}
\pa_{t_{-1}} \eta  = 2  e^{-2\phi} 
\ee
must hold true, 
where $\phi = \phi({x,t_{-1}})$ satisfies 
the sinh-Gordon model \eqref{sg_eq}. Note that
the gauge potential  $A_{t_{-1}}^{\kdv}$ is uniquely determined from
$A_x^{\kdv}$ and the grade structure of the algebra.
Note also that the third term in the gauge potential \eqref{tm11} gives precisely the third
term in the gauge potential \eqref{At_kdv_minus1} thanks to the above identity and the Miura transformation:
\be
\dfrac{1}{4} \left(2 \pa_x \eta \, \pa_{t_{-1}} \eta - \pa_{t_{-1}} \pa_x^2 \eta \right)
= \big( \pa_x \eta + \pa_x^2 \phi - (\pa_x\phi)^ 2\big)e^ {-2\phi}- (\pa_x \phi)^2 e^ {-2\phi} 
= - (\pa_x \phi)^2 e^{-2\phi} .
\ee
Thus, the gauge transformation \eqref{gauge2} automatically maps
the sinh-Gordon  into the negative KdV model \eqref{kdv_neg1}.
Consider now the second negative mKdV flow. 
The gauge transformation \eqref{kdv2}  yields
\be
\label{tm22}
\begin{split}
\widetilde{A}_{\t_{-1}}^{\kdv} & = S_1 A_{t_{-2}}^{\mkdv}  S_1^{-1} + S_1\partial_{t_{-2}} S_1^{-1} \\ 
& =   e^{(\pa_x\phi) E_{-\a}^{(0)}}\left( h^{(-1)}+ 2  e^{-2 \phi} \dI e^{2 \phi} E_{\a}^{(-1)}  
-2 e^{2 \phi} \dI e^{-2 \phi}  E_{-\a}^{(0)} \right) e^{-(\pa_x\phi)E_{-\a}^{(0)}} -\pa_{t_{-2}}\pa_x\phi E_{-\a}^{(0)} 
\\
&= 2  e^{-2 \phi} \big( \dI e^{2 \phi}\big) \left(E_{\a}^{(-1)}+E_{-\a}^{(0)}\right) +\frac{\pa_{t_{-2}\pa_x}\eta}{4} h^{(-1)}+8\left(\pa_x \phi - (\pa_x\phi)^2 e^{-2\phi} \dI e^{2 \phi} \right) E_{-\a}^{(-1)}  ,
\end{split} 
\ee
where we have made use of the Miura transformation \eqref{miura2}.
Since this operator must be unique, i.e., it must be equal to operator \eqref{At_kdv_minus3}, we now conclude
that
\be \label{mt2}
\pa_{t_{-1}}\eta = 4 e^{-2 \phi} \dI e^{2 \phi} ,
\ee
where  
$\phi(x, t_{-2})$ obeys  model \eqref{mkdv_minus_two_eq}.
 Therefore, the gauge transformation yields, besides the standard 
Miura transformation \eqref{miura2}, an additional relation between $\eta_{t_{-2n+1}}$
and one of the two associated mKdV flows, 
$\phi(x, t_{-2n+1})$ or $\phi(x, t_{-2n} )$.
This is the reason why the mapping illustrated in  
diagram~\eqref{corresp_neg} is 2-to-1.
Such additional relations, namely \eqref{mt1} and \eqref{mt2}, 
do not appear when mapping the positive
part of these hierarchies.

We can now generalize the argument for arbitrary negative  flows.
From eq.~\eqref{ker} we know that in general ${\cal D}^{(-1)}_{-N}$ must have the form
 ${\cal D}^{(-1)}_{-N} = \big(f_{-N} \big)\big(E_{-\a}^{(0)} + E_{\a}^{(-1)}\big)$
for some function $f_{-N}$.  
Plugging this into eq.~\eqref{eqofmotionkdv}  and solving for   ${\cal D}^{(-1)}_{-N}$ yield
\be
{\cal D}^{(-1)}_{-N} = \dfrac{1}{2} \big(\pa_{\t_{-N}} \eta\big) \big( E_{-\a}^{(0)} + E_{\a}^{(-1)} \big)  .
\label{Dminus1}
\ee
The gauge transformation \eqref{kdv1} for $N=2n-1$ yields  
a relation between the mKdV field $\phi(x, t_{-2n+1})$ and the KdV field $\eta(x, t_{-2n+1})$:
\be 
\mathcal{D}^{(-1)}_{-N}[\eta] = D^{(-1)}_{-N}[\phi]  
- \big( \pa_{t_{-2n+1}}\pa_x \phi \big)  E_{-\a}^{(0)} .
\label{Dtildeminus1}
\ee
Similarly, the gauge transformation \eqref{kdv2} for $N=2n$ provides a  relation
between $\eta(x, t_{-2n+1})$ and $\phi(x, t_{-2n})$ through
\be
\widetilde {\mathcal{D}}^{(-1)}_{-N}[\eta] = D^{(-1)}_{-N}[\phi]  - \big(\pa_{t_{-2n}}\pa_x \phi \big) E_{-\a}^{(0)} . 
\label{Dtildeminus12}
\ee
Denote $M = 2n-1$ or $M=2n$ and
parametrize 
$D^{(-1)}_{-M}[\phi] = a^{(-1)}_{-M}[\phi] E_{\a}^{(-1)}+ b^{(-1)}_{-M}[\phi] E_{-\a}^{(0)}$,
where $\phi = \phi(x, t_{-M})$.
We  have  from 
the equations of motion of the mKdV hierarchy \eqref{mkdv_neg_eq}  that 
\be
\pa_ {t_{-M}}\pa_x \phi = b^{(-1)}_{-M}- a^{(-1)}_{-M} .
\label{phiba}
\ee
It therefore follows from the above equations 
and   $ {\widetilde {\cal D}}^{(-1)}_{-M} =  {\cal D}^{(-1)}_{-M}$ --- see \eqref{equal_op} --- 
that we must have
\be \label{etaaminus1}
\pa_{\t_{-2n+1}} \eta = 2 a^{(-1)}_{-M} \[ \phi (x, t_{-M}) \]  .
\ee
This implicitly generalizes  the  particular cases of  
identities 
\eqref{mt1} and \eqref{mt2} for all negative  flows:
\be \label{mtgen}
\pa_{\t_{-2n+1}} \eta = 
\begin{cases} 
2 a^{(-1)}_{-2n+1}\[ \phi (x, t_{-2n+1}) \] & \mbox{for odd mKdV flows,} \\  
2 a^{(-1)}_{-2n}\[ \phi (x, t_{-2n}) \] & \mbox{for even mKdV flows.}
\end{cases}
\ee
Naturally, to obtain the explicit form of the function
$a_{-M}^{(-1)}$ one must solve the zero curvature condition grade-by-grade and obtain the
temporal gauge potential explicitly, as previously done for the first two negative flows.
The above argument provides the proof of the correspondence 
summarized in  diagram
\eqref{corresp_pos}.

\subsection{Temporal Miura tranformations}
\label{sec:temporal_miura}

The relations \eqref{mt1} and \eqref{mt2}  are interesting since they allow a mapping
of two mKdV flows into a single KdV flow.
In our previous argument,
the same potential $A_{\t_{-1}}^{\kdv}$ is obtained in two different ways: one by gauging  $A_{t_{-1}}^{\mkdv}$ of  sinh-Gordon,
and the other by gauging $A_{t_{-2}}^{\mkdv}$ of model~\eqref{mkdv_minus_two_eq}. 
By this procedure such relations  are   manifest.  However,
one may still wonder if they are identities or additional conditions.
Let us suppose for the moment that we did not know the underlying algebraic structure of these models nor the gauge
transformations.
Thus, by applying $\pa_{t_{-1}}$ to the Miura transformation \eqref{miura2} (we consider only the minus sign for simplicity) 
and replacing eq.~\eqref{sg_eq} we obtain
\be
\pa_{t_{-1}}\pa_x\eta = 2 \pa_x \phi \, \pa_x\pa_{t_{-1}} \phi - \pa_{t_{-1}}\pa_x^2\phi = -4 \pa_x\phi \, e^{-2 \phi} = 
\pa_x\big( 2 e^{-2\phi} \big) .
\ee
Applying the inverse operator \eqref{dIDef} 
yields $\pa_{t_{-1}} \eta = 2 e^{-2 \phi}$,  i.e.,  precisely
the relation \eqref{mt1}.
The same procedure with \eqref{mkdv_minus_two_eq}  gives instead
\be
\pa_{t_{-1}}\pa_x\eta =  2 \pa_x\phi \, \pa_x\pa_{t_{-2}}\phi - \pa_{t_{-2}}\pa_x^2\phi = 
- 8 \pa_x\phi \, e^{-2\phi} \dI e^{2\phi} + 4 = \pa_x\big( 4 e^{-2\phi} \dI e^{2\phi} \big) ,
\ee
which by the inverse operator \eqref{dIDef} yields 
relation \eqref{mt2}.
These relations are therefore identities, i.e., a consequence
of the Miura transformation and the equations of motion.
In addition, using relations \eqref{mt1} and \eqref{miura2} one can check that 
the differential equation \eqref{kdv_neg1} is identically satisfied, thus establishing 
its correspondence with the sinh-Gordon model. 
The same can be verified 
by replacing relations \eqref{mt2} and \eqref{miura2} 
into the integro-differential equation~\eqref{kdv_neg3} after
tedious manipulations. 
Therefore, such intricate relations could ``in principle''  be derived
 from the Miura transformation plus equations of motion. However, the algebraic construction of these  hierarchies and the gauge transformations establish them directly and 
 systematically.

 \subsection{Dark solitons and peakons}
 \label{sec:solitons_peakons}

We now illustrate  some types of  solutions that can be obtained
from the above connections.
 A powerful approach to construct  solutions of integrable hierarchies 
 is  the \emph{dressing method} \cite{Babelon_1991,Babelon_2003,Ferreira_1997,Hollowod_1993}.
 A crucial ingredient in this approach is 
 a \emph{vertex operator} $F(\k)$ obeying commutator eigenvalue
 equations with the gauge potentials at the vacuum:
 \be
\Big[ 
{A_{t_{N}}}\big\vert_{\text{vac} } \, , F(\k) \Big] = \o_{N}(\k)  F(\k) .
\label{vertex}
\ee
The vertex operator   fixes the dispersion relation\footnote{E.g., a term 
$\xi \equiv \omega_1  x + \omega_{N}  t_N $ in the solution} of any
model within the hierarchy and also  
matrix elements  $A_{ij} \equiv \langle \mu_i | F(\k_1) \dotsm F(\k_n) |\mu_j \rangle$, where $|\mu_i\rangle$ are highest-weight states
of a representation of the Kac-Moody algebra, which completely characterize 
the $n$-soliton interaction terms.
However, under a \emph{nonzero} vacuum configuration a ``deformed'' vertex operator  needs
to be introduced, which  couples the vertex parameter $\k$ with the vacuum background $\v_0$
\cite{Franca_2009}.
Thus, for both zero and nonzero vacuum, 
the dressing approach  yields solutions to the entire hierarchy 
systematically; solutions to different models  
have the same functional form,
the only difference being the dispersion relation.

\begin{figure}[t]
\centering
\includegraphics[width=\textwidth]{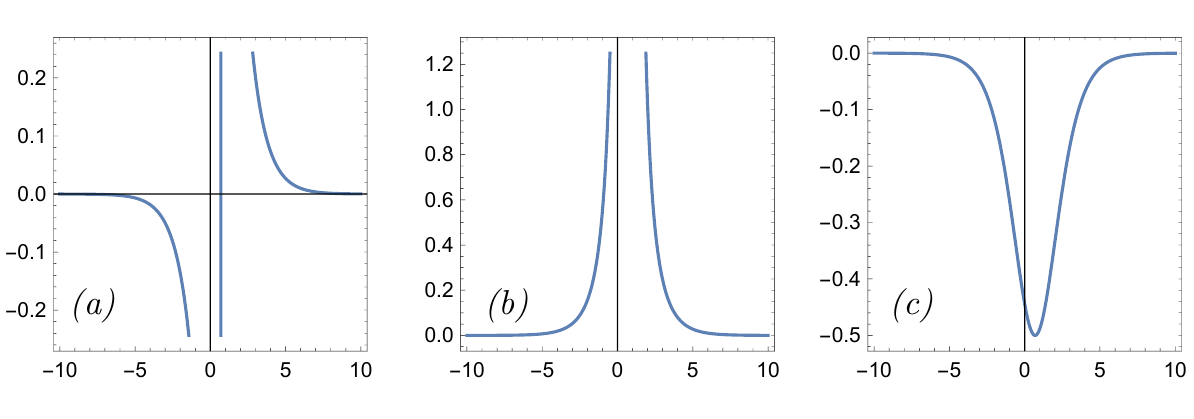}
\caption{A solution of the mKdV hierarchy and its corresponding solutions
in the KdV hierarchy via the Miura transformations \eqref{miura}. \emph{(a)} 
We plot $\v$ given by \eqref{1sol_mkdv} against $\xi$. This is the
equivalent of a 1-soliton solution, but does not
have the usual soliton profile. \emph{(b)} The corresponding solution
of the KdV hierarchy obtained
from  \eqref{miura} with \emph{minus sign}.  This is a \emph{peakon},
which diverges at the mode.
\emph{(c)} A different solution of the KdV hierarchy is obtained from \eqref{miura} with \emph{plus sign}. 
This is a \emph{dark soliton} over a zero vacuum background.
}
\label{plot1}
 \end{figure}

 The \emph{zero vacuum 1-soliton} solution of the mKdV hierarchy
can be constructed from  this approach  yielding \cite{Franca_2009} 
 \be \label{1sol_mkdv}
\v = \pa_x \log \left( \dfrac{\tau_1}{\tau_2} \right), \qquad
\tau_1 = 1 - \dfrac{1}{2} e^{\xi}, \qquad \tau_2 = 1 + \dfrac{1}{2} e^{\xi},
 \ee
where  $\xi$ encodes the dispersion relation of each  model within the hierarchy:
\be\label{disp}
\xi(x,t_N) = 2\k x + 2\k^N t_{N} \qquad
(N = \pm 1, \pm 3, \dotsc).
\ee
Recall that only odd  flows of the mKdV hierarchy admit zero vacuum.
A plot of  \eqref{1sol_mkdv} against $\xi$ is shown
fig.~\ref{plot1}a. Note that $\v \to 0$ as $|x| \to \infty$.
Interestingly, 
starting from a
single solution of the mKdV hierarchy,
the 
Miura transformations~\eqref{miura} induce two  types
of solution of the KdV hierarchy,  one for each sign.
In fig.~\ref{plot1}b we have a so-called \emph{peakon}, which is discontinuous and
diverges at the mode,
 while in fig.~\ref{plot1}c we have a
\emph{dark soliton}.%
\footnote{These solutions are explicitly given by 
$$
\u(x,t_{N}) = \pm \dfrac{16 \kappa^2 e^{2\kappa x + 2\kappa^{N}t_N}}{\big(2\mp e^{2\kappa x + 2k^N t_N}\big)^2} ,
$$
where $N = 2n+1$ for $n=0,\pm 1, \pm 2, \dotsc$.  The first solution is the peakon. The discontinuity comes from the minus sign in the denominator
and differs from the Camassa-Holm equation \cite{Camassa_1993} where the dispersion relation  has an absolute value in the form $| \k x + \omega(t)|$. The second solution flips the sign of the
denominator, yielding the smooth profile of the (dark) soliton.} 
Thus, \emph{all  models within the KdV hierarchy have both peakon and
dark solitons}; they are inherited from the same solution
to the odd models of the mKdV hierarchy. 
The only difference among them is the change
in the dispersion relation~\eqref{disp} which essentially changes the
propagating speed  of such localized waves.
It is also possible to obtain $n$-dark-soliton and $n$-peakon solutions
from the more general solutions of \cite{Franca_2009}.

 \begin{figure}[t]
\centering
\includegraphics[width=\textwidth]{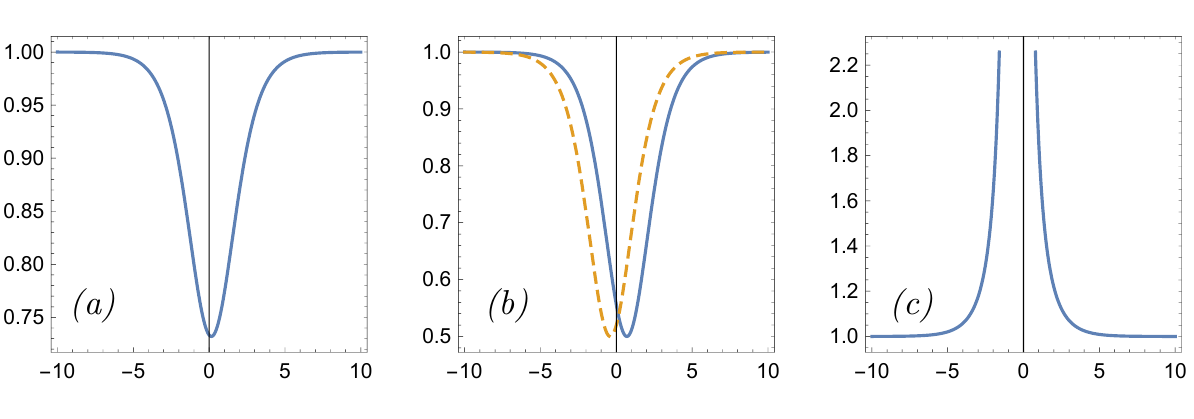}
\caption{\emph{(a)} Dark soliton of the mKdV hierarchy \eqref{nonzerosol}; we set $\v_0=1$ and $\k=1/2$. \emph{(b)} Using the two Miura transformations \eqref{miura} we have dark solitons
of the KdV hierarchy; the solid blue line is with  \emph{plus sign} and the dashed orange
line with  \emph{minus sign}. \emph{(c)} Shifting $\v_0 \to -\v_0$ yields a peakon 
of the KdV hierarchy but now over a nonzero background $\u_0 = \v_0^2$. 
}
\label{plot2}
 \end{figure}

A nonzero vacuum $\v_0$ plays the role of a deformation
parameter in comparison to the affine parameter of the algebra. 
Based on deformed vertex operators \cite{Franca_2009}, the dispersion
relations are obtained from  eq.~\eqref{vertex} and 
 the \emph{nonzero vacuum 1-soliton} of the mKdV hierarchy
reads
\be\label{nonzerosol}
\v = \v_0 + \pa_x \log\left( \dfrac{\tau_1}{\tau_2} \right), \qquad
\tau_1 = 1 + \dfrac{\v_0 - \k}{2 \k} e^{\xi}, \qquad
\tau_2 = 1 + \dfrac{\v_0 + \k}{2 \k} e^{\xi},
\ee
where the dispersion relation couples $\v_0$ and $\k$, e.g.,
for the negative even  flows  we have
\be
\xi(x,t_{-2n}) = 2 \k x + \dfrac{2 \k}{\v_0( \k^2 - \v_0^2)^{n}}  t_{-2n},
\ee
while for the positive part of the hierarchy each dispersion relation
needs to be computed from the respective gauge potential $A^{\mkdv}_{t_{N},\text{vac}}$.\footnote{For instance, for $t_3$ --- mKdV equation --- one finds
$\xi = 2 \k x + (2\k^3 - 3\v_0^2 \k)t_{3}$, while for  $t_{5}$ --- modified Sawada-Kotera ---
one finds $\xi = 2\k x + \big(2\k^5 - 5 \v_0^2 \k + \tfrac{15}{4} \v_0^4 \k\big)t_{5}$, and so on.}
A plot of the solution \eqref{nonzerosol} against $\xi$ 
is shown in fig.~\ref{plot2}a. Note that now we have a
\emph{dark soliton} of the mKdV hierarchy over the constant 
vacuum  $\v \to \v_0$ as $|x| \to \infty$. 
When we replace this solution into the Miura transformations
\eqref{miura}, both signs yield the same type of solution of
the KdV hierarchy --- with just a position shift --- as illustrated in 
fig.~\ref{plot2}b. 
These are again \emph{dark solitons} but now for the KdV hierarchy, with a vacuum $\u_0 = \v_0^2$.
Making $\v_0 \to -\v_0 < 0$ the solution of the 
KdV hierarchy becomes a \emph{peakon over a nonzero background} $\u_0 = \v_0^2$,
as shown in fig.~\ref{plot2}c.
Similarly, $n$-dark-soliton or $n$-peakon solutions over a nontrivial
vacuum can be obtained by plugging in the  more general solutions
proposed in \cite{Franca_2009} into the Miura transformations.

Peakons  were proposed in
the seminal paper \cite{Camassa_1993} through the
Camassa-Holm equation, and later  noted to appear in 
other integrable models  \cite{Degasperis_2002}.
Dark solitons constitute an interesting and active
research topic, with concrete experimental observation
in Bose-Einstein condensates, nonlinear optics, and
condensed matter  physics
\cite{Burger_1999,
Weiner_1988,
Becker_2008,
Delande_2014,
Basnet_2022,
Kopy_2023}.
The above results show that both kinds of solutions are admissible among
the models of the KdV hierarchy, including the KdV equation itself and its
first negative flow \eqref{kdv_neg1}. 
These two different  solutions are obtained from the two possible Miura
transformations leveraging the \emph{same} solution of the 
mKdV hierarchy.
We believe these facts have not been previously noticed in the literature.

\section{Heisenberg subalgebras and commuting flows}
\label{sec:heis}

We have considered individual flows --- differential equations --- of the
mKdV and KdV hierarchies. 
However, an integrable hierarchy must have an infinite number of \emph{mutually commuting
flows}, which are related to an infinite number of involutive conserved charges.
For a \emph{zero vacuum} configuration of the mKdV hierarchy
this is a consequence of the gauge potentials 
having the form $A_{t_N,\vac}^{\mkdv}  = E^{(n)}$ for  $N=2n+1$, 
implying that the operators $A_{t_N,\vac}^{\mkdv}$
are in the kernel of $E$ so  they form     
an \emph{abelian subalgebra} up to a central term, i.e., a Heisenberg subalgebra \cite{Miramontes_1999}.
An important question that we now address is whether this remains true for a \emph{nonzero vacuum} configuration and in particular for 
the \emph{negative even flows} of the mKdV hierarchy. 

Let us first recall some known facts.
Denote the Lax operators of a generic 
integrable  hierarchy by
\be\label{LN}
\mathcal{L}_N \equiv \pa_{t_N} + A_{t_N} ,
\ee
where $\{ t_N \}$ are the admissible time flows (with
$t_1 = x$).
To show that any two given flows commute, 
$\[\partial_{t_N}, \partial_{t_M} \] \mathcal{L}_1$,
it is sufficient
to show that 
$\[ \mathcal{L}_N, \mathcal{L}_{M} \] = 0$;
see, e.g., \cite{Hollowod_1993,Aratyn_2003}.\footnote{Recall that $\mathcal{L}_1  = \pa_x + E + A_0(\{ t_N \})$ is the spatial Lax operator that defines
the hierarchy, with
$E$ being a constant semisimple element and  the fields, 
now depending on all times  $\{ t_N \}$, 
are in the operator
$A_0$.}
For a general field configuration
the Lax operators~\eqref{LN} are related 
to their values at some vacuum via the action
of a dressing operator $\Theta$ \cite{Aratyn_2003}, namely
\be \label{dressing_map}
\mathcal{L}_N  = \Theta \mathcal{L}_{N,\text{vac}}  \Theta^{-1} .
\ee
Therefore
\be \label{ThetaFlow}
\[ \mathcal{L}_N, \mathcal{L}_M \] = \Theta \[ \mathcal{L}_{N, \text{vac}}, \mathcal{L}_{M,\text{vac}} \] \Theta^{-1} 
\ee
and one only needs to show commutation
relations at the vacuum,
$\big[ \mathcal{L}_{N, \text{vac}}, \mathcal{L}_{M,\text{vac}} \big]  = 0$.  
Thus, the infinite set of mutually commuting
flows of an integrable hierarchy, obeying 
zero curvature equations \eqref{ThetaFlow},
is defined with respect to some  vacuum.
Next, we discuss the two relevant cases of interest 
for the purposes of this paper and show that
the mKdV hierarchy can be seen as two ``distinct''
hierarchies depending whether one uses zero 
or nonzero vacuum.

\subsection{Zero vacuum and Type-I mKdV hierarchy}
Let us recall the zero vacuum case $\v \to \v_0 = 0$, which
is well-known  \cite{Miramontes_1999}.
For $N = 2n + 1$ and $M=2m+1$ being \emph{positive odd} or \emph{negative odd}  we have
the Lax operators at the vacuum given by
\be \label{Lmkdvzero}
\mathcal{L}_{N,\vac}^{\mkdvI} = \pa_{t_N} + E^{(N)} ,
\ee
where
\be \label{Ezero}
E^{(N)} \equiv E^{(n)}_\alpha + E^{(n+1)}_{-\alpha} 
= \lambda^n \left( E_\alpha^{(0)} + E_{-\alpha}^{(1)} \right) 
\equiv \lambda^{n} E .
\ee
Note that $E^{(N)} \in \mathcal{K}_E$
which only admits odd-graded elements \eqref{Kernel}.
These operators indeed form an abelian subalgebra\footnote{They form a Heisenberg subalgebra
$\big[E^{(N)}, E^{(M)}\big] = (n-m)\delta_{n+m+1,0}\hat{c}$ if one considers the
central extension, which however plays  no role in defining
the equations of motion of the hierarchy.}
\be \label{abel_zero}
\[  E^{(N)} , E^{(M)}  \] = 0  ,
\ee
from which it follows that $\big[ \mathcal{L}_{N, \text{vac}}^{\mkdvI}, \mathcal{L}_{M,\text{vac}}^{\mkdvI} \big]  = 0$.
Thus,  the zero vacuum configuration
defines an infinite set of mutually commuting flows
indexed by $N, M$ \emph{positive or negative odd}. 
They  form the standard
mKdV hierarchy, now referred to as mKdV \emph{Type-I} for clarity.

Let $\Theta = \Theta_{I}$ be a dressing operator
in eq.~\eqref{dressing_map} for such a zero vacuum configuration.
The gauge-Miura operator  $S$ in eq.~\eqref{gauge} 
maps the Type-I mKdV hierarchy into
the KdV hierarchy, thus
\be \label{gauge_thetaS}
\begin{split}
 \[ \mathcal{L}_N^{\kdv}, \mathcal{L}_M^{\kdv}\] 
 & =  S\[  \mathcal{L}_N^{\mkdvI}, \mathcal{L}_M^{\mkdvI} \]S^{-1} \\ &=
\big( S \Theta_I \big) \[  \mathcal{L}_{N,\vac}^{\mkdvI}, \mathcal{L}_{M,\vac}^{\mkdvI} \]\big( S \Theta_I \big) ^{-1} \\ &= 0 ,
\end{split}
\ee
showing that the flows of the KdV hierarchy in
the orbit of such a zero vacuum  commute.

\subsection{Nonzero vacuum and Type-II mKdV hierarchy}
We now consider a constant, nonzero, vacuum $\v \to \v_0\ne 0$.  
By careful inspection of the Lax operators
defining the equations of motion of the mKdV hierarchy
from Secs.~\ref{sec:pos_flow} and \ref{sec:mkdv_neg}
we conclude that they have the form
\begin{subequations} \label{Lnonzero}
\begin{align}
\mathcal{L}^{\mkdvII}_{1,\vac} &= \pa_{x} + \bb , \label{Lvac1} \\
\mathcal{L}_{-2n,\vac}^{\mkdvII} &= \pa_{t_{-2n}} + (\v_0)^{-1} \bb^{(-2n+1)}, \label{Lvaceven} \\
\mathcal{L}_{2n+1, \vac}^{\mkdvII} &= \pa_{t_{2n+1}} + \sum_{j=1}^n c_j \v_0^{2n-2j} \bb^{(2j+1)} ,
\label{Lvacodd}
\end{align}
\end{subequations}
for some numbers $c_j$ in \eqref{Lvacodd} and where ($N= 2n+1$, $n=0,\pm1,\dotsc$)
\be \label{Bdef}
\bb^{(N)} \equiv  E^{(N)} + \v_0 h^{(n)} = \lambda^{n} (E + \v_0 h) \equiv\lambda^n \bb .
\ee
Note that $\bb$ contains the vacuum $\v_0$ as a parameter.
The term $E^{(N)}$ has degree $2n+1$ and $h^{(n)}$ has degree
$2n$ according to the principal gradation $\widehat{Q} \equiv 2 \widehat{d} + \tfrac{1}{2} h^{(0)}$.
If we associate degree $1$ to $\v_0$, i.e., if we redefine the grading
operator as $\widetilde{Q} \equiv \widehat{Q} + \widetilde{d}$, where
$\widetilde{d} \equiv \v_0 \tfrac{\pa}{\pa \v_0}$, then
$\bb^{(N)}$ has  degree  $2n+1$ 
and it is a sum of two \emph{homogeneous} terms of degree
$2n+1$. Thus $\v_0$ can be interpreted as a new spectral
parameter defining a two-loop algebra as discussed
in \cite{Aratyn_1992,Schwimmer_1992}.\footnote{%
The two-loop algebra is given by 
$$
\[ T^{(m,r)}_a,  T_b^{(n,s)} \] = i f_{ab}^c T^{(m+n, r+s)}_c  + 
\widehat{c} \delta_{ab} r \delta_{r+s, 0} \delta_{m+n, 0} + 
\widetilde{c} \delta_{ab} m \delta_{m+n,0} \delta_{r+s,0} ,
$$
where $\widehat{c}$ and $\widetilde{c}$  are two central terms,
and $\widehat{d}$ and $\widetilde{d}$ are  two derivation operators such that
$\big[ \widehat{d}, T^{(m,r)} \big] = m T^{(m,r)}$ and    
$\big[ \widetilde{d}, T^{(m,r)}\big] = r T^{(m,r)}$.
The generators can be  realized  as 
$T_a^{(m,n)} = T_a \lambda^{m} \gamma^{r}$, 
$\widehat{d}= \lambda \tfrac{\pa }{\pa \lambda}$, 
and $\widetilde{d} = \gamma \tfrac{\pa}{\pa \gamma}$. 
Thus, $\v_0 \sim \gamma$ plays the role of the second spectral parameter.
}
Each individual term in the sum \eqref{Lvacodd} has  degree $2n + 1$, which is also the highest
degree of the operator $\mathcal{L}_{2n+1}^{\mkdv}$.
Similarly, the operator \eqref{Lvaceven} has degree $-2n$,
which is the lowest degree of operator $\mathcal{L}_{-2n}^{\mkdv}$.
From \eqref{Bdef} we again have
an \emph{abelian subalgebra},
\be
\[ \bb^{(N)}, \bb^{(M)}\] = 0,
\ee
in close analogy with the zero vacuum 
case \eqref{abel_zero}.\footnote{We also have the ``deformed'' Heisenberg
subalgebra
$\big[\bb^{(N)}, \bb^{(M)}\big] = \big( (n-m)\delta_{n+m+1,0}
+ 2 \v_0^2 n \delta_{n+m,0}\big) \hat{c}$ under the
central extension.
}
This implies that for $N,  M$  indexed as  in eqs.~\eqref{Lnonzero}, i.e., \emph{positive odd or  negative even}, 
$\big[ \mathcal{L}_{N,\vac}^{\mkdvII} , \mathcal{L}_{M,\vac}^{\mkdvII}\big] = 0$.
By the dressing transformation \eqref{ThetaFlow}
with a suitable  operator $\Theta = \Theta_{II}$,
we have  
$\big[ \mathcal{L}_N^{\mkdvII}, \mathcal{L}_M^{\mkdvII}\big] = 0$ 
for a general field configuration
in the orbit of such a nonzero vacuum --- which is parametrized by $\v_0$.
Therefore, such mutually commuting flows define a proper integrable hierarchy,
which we refer to as mKdV \emph{Type-II}.
We provide an explicit example of such commuting flows
in sec.~\ref{sec:examplecomm} in the appendix.

Thus, the mKdV-II hierarchy has positive odd and negative even 
 flows that commute among themselves.
The gauge-Miura transformation \eqref{gauge2} and \eqref{kdv2}
also maps the mKdV-II hierarchy into the KdV hierarchy.
More precisely, 
under the dressing transformation \eqref{dressing_map} we now have
\be \label{gauge_thetaS2}
\begin{split}
 \[ \mathcal{L}_{\lceil N \rceil} ^{\kdv}, \mathcal{L}_{\lceil M \rceil}^{\kdv}\] 
& =  S\[  \mathcal{L}_N^{\mkdvII}, \mathcal{L}_M^{\mkdvII} \]S^{-1} \\
& =
\big( S \Theta_{II} \big) \[  \mathcal{L}_{N,\vac}^{\mkdvII}, \mathcal{L}_{M,\vac}^{\mkdvII} \]\big( S \Theta_{II} \big) ^{-1} \\ 
&= 0  ,
\end{split}
\ee
showing that the flows of the KdV hierarchy in the orbit of such a nonzero
vacuum --- parametrized by  $\u_0 = \v_0^2$  --- also commute.
For consistency of notation, above 
we defined $\lceil N \rceil \equiv N$ for $N=2n+1$ (positive odd) and
$\lceil N \rceil \equiv -2n+1$ for $N = -2n$  (negative even).

By analogous reason that the operators \eqref{Lmkdvzero} only admit
odd  flows, 
the operators \eqref{Lnonzero} only admit negative
even flows, besides the positive odd ones. 
The reason is the following.
For negative even flows ($N = -2n$), 
$A_{t_N,\vac}^{\mkdv} = \big( D^{(-2n)}_\vac + D^{(-2n+1)}_\vac\big) + \dotsm + \big( D^{(-2)}_\vac + D^{(-1)}_\vac \big)$, where each
term in parenthesis combines precisely into a term proportional to $\bb^{(j)}$,
i.e., $D^{(2j)}_\vac + D^{(2j-1)}_\vac \sim \bb^{(j)}$ --- 
actually, only the first term survives yielding the
form \eqref{Lvaceven}.
The same structure repeats itself for positive odd flows ($N = 2n+1$), namely
$A_{t_{N},\vac}^{\mkdv} = 
\big( D^{(2n+1)}_\vac + D^{(2n)}_\vac\big)
+ \dotsm +
\big( D^{(1)}_\vac + D^{(0)}_\vac\big)
$ --- here all terms survive yielding the form \eqref{Lvacodd}.
However, for negative odd flows ($N = -2n + 1$)
$A_{t_N,\vac}^{\mkdv} =  
D^{(-2n+1)}_\vac +  \dotsm + D^{(-1)}_\vac$
contains an odd number of terms and therefore can never combine
into a sum of  $\bb^{(j)}$'s  alone.
Therefore:
\begin{itemize} 
\item  The mKdV-I hierarchy has \emph{positive odd} and 
\emph{negative odd} flows, and  is defined
in the orbit of a \emph{zero vacuum} ($\v_0=0$); 
\item The mKdV-II hierarchy has \emph{negative even} and \emph{positive odd}  flows, and is defined
in the orbit of a \emph{nonzero vacuum} (parametrized by $\v_0 \ne 0$);
\item Notably,
the differential equations for the negative parts of mKdV-I and mKdV-II are  different.
However, both mKdV-I and mKdV-II have the same differential
equations for their positive parts.  
Such positive flow equations therefore allow both
types of solution, i.e., with zero (mKdV-I) and nonzero (mKdV-II) vacuum.
\item The flows within each mKdV-I and mKdV-II commute among
themselves, but crossed flows between them do not. 
That is, only flows defined in the orbit of the same vacuum commute.
\end{itemize}
Furthermore, the same gauge-Miura transformation $S$ maps
both mKdV-I and mKdV-II  into the
KdV hierarchy that has only \emph{odd flows} (positive and negative);
see eqs.~\eqref{gauge_thetaS} and \eqref{gauge_thetaS2}.\footnote{In the same vein,
we have the ``KdV-I'' hierarchy of mutually commuting flows that  is
defined in the orbit of a zero  vacuum,  and the 
``KdV-II'' hierarchy defined in the orbit of a nonzero vacuum. 
Note,  however, that both KdV-I and KdV-II have exactly the same
differential equations, so we  refer to them simply as KdV, 
in contrast   to mKdV-I and mKdV-II.}
This explains why the entire KdV hierarchy admits both types of solution, i.e.,
with zero and nonzero vacuum. 
In light of this discussion, the 
diagrams \eqref{corresp_pos} and \eqref{corresp_neg} should be seen
as a shorthand for
\be
\begin{tikzcd}[row sep=0.2em, column sep=2em]
t_{N}^{\mkdvI} \arrow[r, "S"] & t_{N}^{\kdv} 
\end{tikzcd} \qquad \text{and}
\qquad
\begin{tikzcd}[row sep=0.2em, column sep=2em]
t_{M}^{\mkdvII} \arrow[r, "S"] & t_{\lceil M \rceil}^{\kdv} 
\end{tikzcd} 
\ee
where $N$ is positive or negative odd, and $M$ is positive odd or negative even.
Note also that commuting flows of the KdV hierarchy are  defined
in the orbit of some vacuum, defined by the mappings 
\eqref{gauge_thetaS} and \eqref{gauge_thetaS2}. However, contrary to 
mKdV, the differential
equations in the KdV hierarchy  
are the same in both situations.

\section{Extension to $\widehat{s\ell}(3)$}
\label{sec:extension}

The previous ideas  generalize to more 
complex cases such as integrable models constructed from   
$\widehat{s\ell}(3)$ or more generally from $\widehat{s\ell}(r+1)$; see the appendix~\ref{sec:algebra} for definitions.
Let us consider explicitly the $\widehat{s\ell}(3)$ case first.
The analog of the mKdV gauge potential \eqref{Ax_mkdv} has now two fields, $\vv(x,t_N)$ and 
$\vb(x,t_N)$, and
assumes the form
\be\label{Ax_mkdvsl3}
A_x^{\mkdv} 
= E + A_{0}
= E_{\alpha_1}^{(0)}
+ E_{\alpha_2}^{(0)}
+E_{-\alpha_1-\alpha_2}^{(1)}+  \vv  h_1^{(0)} + \vb  h_2^{(0)} .
\ee
Similarly, the KdV gauge potential \eqref{Ax_kdv} acquires two fields, $\uu(x,t_{N})$ and $\ub(x,t_N)$, 
where
one of them is with $A_{-1} \in \lie_{-1}$  and the other is 
with $A_{-2} \in \lie_{-2}$,%
\footnote{%
Matrix representations for \eqref{Ax_mkdvsl3} and 
\eqref{Ax_kdvsl3} are
$\left( \begin{smallmatrix}
 \vv & 1 & 0 \\
 0 & \vb-\vv & 1 \\
 \lambda  & 0 & -\vb \\
\end{smallmatrix}\right)$
and
$\left(\begin{smallmatrix}
 0 & 1 & 0 \\
  0 & 0 & 1 \\
 \ub + \l & \uu & 0
\end{smallmatrix} \right) $,   
respectively.}
\be\label{Ax_kdvsl3}
A_x^{\kdv} = E + A_{-1} + A_{-2}
= E_{\a_1}^{(0)} + E_{\a_2}^{(0)} + E_{-\a_1-\a_2}^{(1)} + 
\uu  E_{-\a_1}^{(0)} + \ub  E_{-\a_1 - \a_2}^{(0)} .
\ee
Instead of two gauge-Miura transformations \eqref{gauge}  now there are three, $S \in \{S_1, S_2, S_3\}$, all leading
to the same correspondence (see \cite{Lobo_2021} for details).%
\footnote{The number of gauge transformations, yielding
the number of possible Miura transformations, is the rank $r  + 1$   of
the affine Lie algebra (see the appendix of \cite{Olive_1985}) and each $S$ is directly associated with an element of the kernel of $E$ (whose order is given by the exponent of $\widehat{s\ell}(r+1)$).} 
We show results for 
one of them for simplicity, namely
\be \label{gaugesl3}
A_{x}^{\kdv}  =  S_1 A_{x}^{\mkdv} S_1^{-1} + S_1\partial_{x} S_1^{-1}   ,
\ee
where
\be
S_1 =  e^{\g_3 E_{-\a_1-\a_2 }^{(0)}}e^{\g_2 E_{-\a_2 }^{(0)}}e^{\g_1 E_{-\a_1}^{(0)}} 
=   
\begin{pmatrix}
1 & 0 & 0 \\ 
\pa_x\phii & 1 & 0 \\ 
(\pa_x \phii)^2 - \pa_x^2\phii & \pa_x\phib & 1 \\ 
\end{pmatrix} ,
\label{gsl3}
\ee
 $ \vv \equiv \pa_x \phii$,  $ \vb \equiv \pa_x \phib$, 
$\g_1= \pa_x\phii$, 
$\g_2 = \pa_x\phib$, 
and $\g_3 = (\pa_x\phii)^2 - (\pa_x\phii)( \pa_x \phib) - \pa_x^2\phii$.
Such a gauge transformation realizes a Miura-type  transformation among the fields: 
\begin{subequations}
\label{miurasl3}
\begin{align}
\uu &=  - \pa_x\vv - \pa_x\vb + \vv^2 - \vv \vb + \vb^2, \\
\ub &= - 2 \vv \pa_x \vv + \vv \pa_x \vb + \pa_x^2\vv + \vv^2 \vb  - \vv \vb^2. 
\end{align}
\end{subequations}

As before, such a gauge transformation provides a  1-to-1 correspondence between the models of the positive part of these hierarchies. 
However, the negative part is more subtle and now
a \emph{triple} of temporal mKdV gauge potentials, 
 $A_{t_{-3n+2}}^{\mkdv}$, $A_{t_{-3n+1}}^{\mkdv}$  and $A_{t_{-3n}}^{\mkdv} $, 
fuse into a \emph{single}
temporal KdV gauge potential, $A_{t_{-3n + 2}}^{\kdv}$,
as summarized  by the diagram 
\be\label{corresp3}
\begin{tikzcd}[row sep=0.4em, column sep=3em]
t_{N}^{\mkdv} \arrow[r, "S"] & t_{N}^{\kdv} \\
\\
t_{-N}^{\mkdv} \arrow[dr, "S"] & \\
t_{-N-1}^{\mkdv} \arrow[r, "S"] & t_{-N}^{\kdv} \\ 
t_{-N-2}^{\mkdv} \arrow[ur, swap, "S"]\\
\end{tikzcd} 
\ee
For the positive part the index $N$ can 
assume values $3n+1$ or $3n+2$, for $n=0, 1,\dotsc$. This is because
the highest grade component of the zero curvature condition, 
$\big[ E, \mathcal{D}^{(N)}_{N} \big] = 0$,
implies that $\mathcal{D}^{N}_N \in \mathcal{K}_E$, which from
eq.~\eqref{Kernelsl3} there are two possibilities,
each leading to a different model.
However, for the negative part we have
$- N = 3n + 2$, for $n=-1,-2,\dotsc$.\footnote{Here  the first lowest
grade component of the zero curvature condition is
$\big[ A_{-2}, \mathcal{D}^{(-M)}_{-N} \big]  = 0$,  implying
that $\mathcal{D}^{-M}_{-N} \sim E^{(m)}_{-\alpha_1-\alpha_2}$ for $m=0,-1,\dotsc$. 
According to \eqref{sl3grade} such an element has grade
$3m+1$, hence 
$-M = 3m+1 = -2, -5, \dotsc$.
Adopting the \emph{convention} that the first negative flow is labelled as $t_{-1}$, in order to
match the mKdV terminology, for the negative part we denote
$-N \equiv -M + 4$ so that $-N = -1,-4,\dotsc = 3n +2$ for $n=-1,-2,\dotsc$.
Now the lowest grade term in $A_{t_{-N}}$ is $\mathcal{D}^{-M}_{-N} = \mathcal{D}^{-N-4}_{-N}$.}

The derivation of the equations of motion from the zero
curvature condition, and
the explicit gauge transformation for the temporal gauge potentials,
follow exactly the same procedure as previously explained 
for $\widehat{s\ell}(2)$ although the
calculations are longer. 
We thus limit the discussion to stating the main results for the sake of simplicity.
Some of the representative models within these hierarchies, related by the above diagram, are  described as follows.

\begin{itemize}

\item \underline{$t_{1}$-flow} \ 
This case gives the chiral wave equations
$\pa_{t_1} \vv = \pa_x \vv$ and $\pa_{t_1} \vb = \pa_x \vb$
for the 2-component mKdV, and similar equations for the 2-component
KdV.

\item \underline{$t_{2}$-flow} \
For the  $\widehat{s\ell}(3)$-mKdV  hierarchy we have the equations of motion
\begin{subequations}
\label{mkdvt2sl3}
\begin{align}
3  \pa_{t_2}\vv  &=  \pa_x \left( \vv^2  + 2 \vv \vb  - 2 \vb^2 - \pa_x \vv  + 
2 \pa_x \vb  \right),  \\
3 \pa_{t_2} \vb  &=  \pa_x \left( 2 \vv^2 - 2 \vv \vb - \vb^2 
-2 \pa_x \vv+ \pa_x \vb \right) .
\end{align}
\end{subequations}
Its counterpart in the $\widehat{s\ell}(3)$-KdV hierarchy is given by
\begin{subequations}
\label{kdvt2sl3}
\begin{align}
\pa_{t_2} \uu  &= \pa_x \left(  2 \ub + \pa_x \uu   \right),  \\
3 \pa_{t_2} \ub &=  \pa_x \left( \uu^2  - 3\pa_x \ub  - 2 \pa_x^2 \uu   \right).
\end{align}
\end{subequations}

\item \underline{$t_{-1}$-flow (mKdV)}  \
The $\widehat{s\ell}(3)$-mKdV hierarchy yields an affine Toda field
theory as the first negative flow given by 
\begin{subequations}
\label{mkdvsl3_minus1}
\begin{align}
\pa_{t_{-1}} \pa_{x} \phii &= e^{2\phii - \phib}-e^{-\phii-\phib}, \\
\pa_{t_{-1}} \pa_{x} \phib &= e^{-\phii + 2\phib }-e^{-\phii - \phib} ,
\end{align}
\end{subequations}
where $\vv \equiv \pa_x \phii$ and $\vb \equiv \pa_x \phib$.
\item \underline{$t_{-2}$-flow (mKdV)} We have the integro-differential
equations
\begin{subequations} \label{mkdvsl3_minus2}
\begin{align}
\pa_{t_{-2}} \pa_{x} \phii &= e^{ 2\phii -\phib} \dI \left[ e^{-\phii-\phib} 
 - e^{-\phii + 2 \phib  }\right]  \nonu  \\
&  - e^{-\phii-\phib} \dI \left[ e^{-\phii+2\phib} - e^{2\phii-\phib}\right] , 
\\
\pa_{t_{-2}} \pa_{x} \phib &=-e^{-\phii+2\phib} \dI \left[e^{-\phii-\phib} 
- e^{2\phii-\phib}\right] \nonu \\
&  - e^{-\phii-\phib} \dI \left[ e^{-\phii + 2\phib}- e^{2\phii-\phib}\right] .
\end{align}
\end{subequations}

\item \underline{$t_{-3}$-flow (mKdV)} In this case we obtain
\begin{subequations}
\label{mkdvsl3_minus3}
\begin{align}
\pa_{t_{-3}}\pa_{x} \phii &= 
e^{\phii-\phib} \dI \left[  e^{ 2\phii-\phib}  \dI e^{ -\phii + 2\phib } \right] 
\nonu \\ 
&  -e^{\phii-\phib} \dI \left[ e^{-\phii + 2\phib } \dI e^{ 2\phii-\phib}\right] \nonu \\
&  + 2 e^{2\phii-\phib} \dI \left[ e^{ -\phii + 2\phib}  \dI e^{ -\phii-\phib}\right] \nonu \\
&  + e^{2\phii-\phib} \dI \left[ e^{ -\phii-\phib} \dI e^{ -\phii+2\phib} \right], \\
\pa_{t_{-3}}\pa_{x} \phib &=
 -e^{\phii-\phib} \dI \left[ e^{ -\phii+2\phib}  \dI  e^{ 2\phii-\phib}\right] 
 \nonu \\
&   +e^{\phii-\phib} \dI \left[ e^{2\phii-\phib} \dI e^{ -\phii+2\phib}\right] \nonu \\
& -2 e^{ -\phii + 2 \phib  } \dI \left[ e^{ 2\phii-\phib}  \dI e^{ \phii-\phib} \right] \nonu \\ 
&  -e^{-\phii + 2\phib } \dI \left[ e^{ -\phii-\phib} \dI e^{ 2\phii-\phib}\right] .
\end{align}
\end{subequations}

\item \underline{$t_{-1}$-flow (KdV)} For the first negative flow
of the $\widehat{s\ell}(3)$-KdV hierarchy we obtain
\begin{subequations}\label{kdvsl3_minus1}
\begin{align}
%
0 &=  2 \pa_{t_{-1}}  \pa_x\etaa \, (\pa_x\etaa)^2 
+ 2 \pa_{t_{-1}}\etaa \, \pa_x^2 \etaa \, \pa_x \etaa 
-4 \pa_{t_{-1}} \pa_{x}^3  \etaa \, \pa_x \etaa \nonu \\
&  -3 \pa_{t_{-1}} \pa_{x}^2  \etab  \, \pa_x \etaa   
  - 6  \pa_x^2 \etaa \, \pa_{t_{-1}} \pa_x^2 \etaa  
- 6 \pa_{t_{-1}} \pa_x  \etaa \, \pa_{x}^3  \etaa  
- 2 \pa_{t_{-1}} \etaa \, \pa_x^4 \etaa   \nonu \\
&  + 2  \pa_{t_{-1}} \pa_x^5  \etaa  
  + 9 \pa_x \etab \, \pa_{t_{-1}}\pa_x \etab  
 - 6  \pa_{t_{-1}} \pa_x  \etaa \, \pa_x^2  \etab 
+ 3 \pa_{t_{-1}} \etab \, \pa_x^2 \etab  \nonu \\
&  - 3 \pa_{t_{-1}} \etaa \, \pa_x^3 \etab
+ 3 \pa_{t_{-1}} \pa_x^4 \etab , \\
0 &= 2  \pa_{t_{-1}} \pa_{x}  \etaa \, \pa_{x}^2 \etaa + \pa_{t_{-1}} \etab \, \pa_x^2 \etaa 
+ \pa_x \etaa \, \pa_{t_{-1}}  \pa_x^2 \etaa  + \pa_{t_{-1}} \etaa \, \pa_x^3 \etaa   \nonu \\ 
& -   \pa_{t_{-1}} \pa_x^4  \etaa  
  + 3 \pa_{t_{-1}} \pa_x \etaa  \, \pa_x \etab
+ 2  \pa_x \etaa \, \pa_{t_{-1}}  \pa_x \etab \nonu \\
& + 2  \pa_{t_{-1}}\etaa \,  \pa_x^2 \etab - 2 \pa_{t_{-1}} \pa_x^3  \etab ,
\end{align}
\end{subequations}
where   $ \uu \equiv \pa_x \etaa  $ and $ \ub \equiv \pa_x \etab $.
The above model corresponds to the counterpart of the affine Toda field theory \eqref{mkdvsl3_minus1} in the $\widehat{s\ell}(3)$-KdV hierarchy.
Actually, 
all three mKdV models \eqref{mkdvsl3_minus1}, \eqref{mkdvsl3_minus2} and
\eqref{mkdvsl3_minus3} are connected to model \eqref{kdvsl3_minus1} 
by gauge transformation; see diagram \eqref{corresp3}.

\item Higher positive flows, or lower negative ones, for both hierarchies
can be obtained as well, although the equations quickly become
 quite complicated.

\end{itemize}

For the affine Toda model \eqref{mkdvsl3_minus1} we have, besides the Miura-type transformation \eqref{miurasl3}, 
the  ``temporal Miura relations''
\be \label{sl3_temp_miura}
\pa_{t_{-1}} \etaa =  3 e^{-\phii-\phib}, \qquad 
\pa_{t_{-1}} \etab =  3 (\pa_x \phii)  e^{-\phii-\phib}  ,
\ee
which are the analog of relation \eqref{mt2} for sinh-Gordon.
The map between
models \eqref{mkdvsl3_minus2} and \eqref{kdvsl3_minus1} involve
instead the relations
\begin{subequations}
\begin{align}
\pa_{t_{-1}} \etaa  &=  3 e^{-\phii-\phib} \dI  \left( e^{-\phii + 2 \phib }
- e^{2 \phii-\phib}\right), \\ 
\pa_{t_{-1}} \etab &= 3 e^{\phii-2\phib}+ 3 e^{-\phii-\phib}  
(\pa_x \phii)  \dI \left( e^{-\phii + 2 \phib}- e^{2 \phii-\phib}\right),
\end{align}
\end{subequations}
and in the case of model \eqref{mkdvsl3_minus3} we have
\begin{subequations}
\begin{align}
\pa_{t_{-1}} \etaa &= 
3 e^{-\phii - \phib} \dI\left( 
e^{-\phii + 2 \phib }  \dI  e^{2 \phii - \phib}
- e^{2 \phii - \phib}  \dI 
e^{-\phii + 2 \phib } \right) , \\
\pa_{t_{-1}} \etab &= 3  e^{\phii - 2 \phib} 
\dI e^{-\phii + 2 \phib}   \nonu \\
 & + 3 (\pa_x \phii ) e^{-\phii-\phib} 
 \dI 
\left[ e^{-\phii + 2 \phib } \dI e^{2 \phii -
\phib} - e^{2 \phii -\phib} \dI  e^{-\phii  + 2 \phib }
\right]. 
\end{align}
\end{subequations}
Such relations follow immediately from the zero curvature construction and
 the gauge-Miura transformations; they  would otherwise be impossible to guess, i.e., based on 
the equations of motion  and the Miura transformation \eqref{miurasl3} alone.

The vacuum structure for the negative flows follow a similar structure as the previous
$\widehat{s\ell}(2)$ case, and will be discussed in more generality in 
sec.~\ref{sec:extension2}.

\subsection{Tzitz\' eica-Bullough-Dodd and its KdV counterpart}
\label{sec:BD}

As a particular case of the affine Toda  theory \eqref{mkdvsl3_minus1}, setting  $\phi_1 = \phi_2 = -\varphi$
we obtain 
the Tzitz\' eica-Bullough-Dodd model in light cone coordinates:
\be\label{BD_model}
\pa_{t_{-1}} \pa_x \varphi = e^{2\varphi} - e^{-\varphi}. 
\ee
Form factors of this model were
obtained in \cite{Fring_1993}, and
its relation with the 
Izergin-Korepin massive quantum field theory through
Bethe ansatz equations was demonstrated in
\cite{Dorey_2013}, as well as its conformal limit.
Under this reduction, the Miura transformation
\eqref{miurasl3} implies
\begin{subequations}
\begin{align}
\uu &\equiv \pa_x \eta  = (\pa_x \varphi)^2 +  2 \pa_x^2 \varphi , \\
\ub &\equiv \pa_x \eta_2 = - \tfrac{1}{2} \pa_x^2 \eta,
\end{align}
\end{subequations}
so that  system \eqref{kdvsl3_minus1} reduces to
\be \label{bd_kdv}
\begin{split}
     2& \pa_{t_{-1}} \pa_{x}^5\eta +8 \big( \pa_{t_{-1}} \pa_{x}\etaa\big) \big(\pa_{x}\eta\big)^2+ 2 \big(4 \pa_{t_{-1}}\eta \, \pa_{x}^2 \eta
     -5\pa_{t_{-1}} \pa_{x}^3\eta \big) \big(  \pa_{x}\eta  \big) \\ & \qquad 
     -15 \pa_{x}^2 \eta \, \pa_{t_{-1}} \pa_{x}^2\eta 
     -9 \pa_{t_{-1}} \pa_{x}\eta \, \pa_{x}^3\eta -2 \pa_{t_{-1}}\eta \, \pa_{x}^4\eta =0 .
\end{split}
 \ee
 Moreover, the temporal Miura relation \eqref{sl3_temp_miura} yields
\be
\pa_{t_{-1}} \eta = 3 e^{2\varphi}.
\ee
 The  integrable model \eqref{bd_kdv}
 corresponds to the KdV counterpart of
the Tzitz\'eica-Bullough-Dodd   \eqref{BD_model}, in the same way that the
 sinh-Gordon  is related to the first negative
 KdV flow \eqref{kdv_neg1}.
 Both   sinh-Gordon and  Tzitz\' eica-Bullough-Dood can be seen
 as (different) integrable perturbations of the conformal
 Liouville  theory, which is obtained from Einsteins's
 equations in 2D and has important connections
 in  string theory  \cite{Polyakov_1981}.

\subsection{Commuting flows}
\label{sec:commflow_sl3}

The same arguments used for  $\widehat{s\ell}(2)$ ---
see  sec.~\ref{sec:heis} ---  
extend to $\widehat{s\ell}(3)$ as we now show. 
The spatial Lax operator is
\be
\begin{split}
\mathcal{L}_1^{\mkdv} &= \pa_x + E + A_0 \\
&= \pa_x +  E_{\alpha_1}^{(0)} + E_{\alpha_2}^{(0)} + E_{-(\alpha_1+\alpha_2)}^{(1)}
+ \v_1  h_1^{(0)}+ \v_2  h_2^{(0)} .
\end{split}
\ee
The kernel of $E$ (recall sec.~\ref{sec:sl3} in the appendix) has elements 
$\mathcal{K}_E = \{ E^{(3n+1)} ,  E^{(3n+2)}  \}$  with
\begin{subequations}
\begin{align}
E^{(3n+1)}  &\equiv E_{\alpha_1}^{n} + E_{\alpha_2}^{(n)} + E_{-\alpha_1-\alpha_2}^{(n+1)} = \lambda^{n} E, \\
E^{(3n+2)} &\equiv E_{-\alpha_1}^{(n+1)} + E_{-\alpha_2}^{(n+1)} + E_{\alpha_1 + \alpha_2}^{(n)} = \lambda^{n+1} E^{\dagger} .
\end{align}
\end{subequations}
For \emph{zero vacuum} configuration $\v_i \to \v_{i,0} = 0$ ($i=1,2$)
we have
\begin{subequations}
\begin{align}
\mathcal{L}^{\mkdvI}_{3n+1,\vac}  &= 
\pa_{t_{3n+1}} + E^{(3n+1)}  , \\
\mathcal{L}^{\mkdvI}_{3n+2,\vac} &= \pa_{t_{3n+2}} + E^{(3n+2)} ,
\end{align}
\end{subequations}
for $n=0,\pm 1, \pm 2, \dotsc$.
These include both the positive and  negative flows
that admit zero vacuum. As for the $\widehat{s\ell}(2)$ case \eqref{abel_zero}, we also have an
abelian subalgebra up to a central term:
\be \label{commEE}
\left[ E^{(3n+p)}, E^{(3m+p')}  \right] = 0 \qquad (p,p' = 1,2) .
\ee
This implies commuting flows at the vacuum and, as a consequence of 
 dressing transformations, also
$\big[ \mathcal{L}_{3n+p}^{\mkdvI}, \mathcal{L}_{3m+p'}^{\mkdvI} \big] = 0$
for a general field configuration in its orbit.
By gauge-Miura \eqref{gaugesl3} this implies
commuting flows of the KdV hierarchy, analogously to
\eqref{gauge_thetaS}. Note, however, that the mapping is 
degenerate, i.e., there is a 2-fold mapping
for the negative part of the hierarchies described in the
diagram \eqref{corresp3} regarding  zero vacuum, i.e.,
this relation can  be broken down into
\be
\begin{tikzcd}[row sep=0.2em, column sep=2em]
t_{3n+1}^{\mkdvI} \arrow[r, "S"] & t_{3n+1}^{\kdv} \\
t_{3n+2}^{\mkdvI} \arrow[r, "S"] & t_{3n+2}^{\kdv} 
\end{tikzcd} \qquad \text{and}
\qquad
\begin{tikzcd}[row sep=0.0em, column sep=2em]
t_{-3m+1}^{\mkdvI} \arrow[dr, "S"] & \\
&  t_{-3m+2}^{\kdv} \\ 
t_{-3m+2}^{\mkdvI} \arrow[ur, swap, "S"] 
\end{tikzcd} 
\ee
where $n=0,1,2,\dotsc$ (left diagram) accounts for  positive
flows and  $m=1,2,\dotsc$ (right diagram) accounts for 
negative flows. 

Moving on to the \emph{nonzero vacuum} case
$\v_i \to \v_{i,0} \ne 0$ ($i=1,2$), we have
the Lax operator
\be \label{BB}
\mathcal{L}_{1,\vac}^{\mkdvII} 
= \pa_x + \bb, \qquad \bb \equiv E + \v_{1,0} h_1^{(0)} + \v_{2,0} h_2^{(0)} .
\ee
It turns out that the element $\bb$ has a well-defined kernel 
\be \label{kernelB}
\mathcal{K}_{\bb} = \{ \bb^{(3n+1)} , \bb^{(3n+2)}  \} ,
\ee
where
\begin{subequations} \label{kerBelems}
\begin{align}
\bb^{(3n+1)} &= E^{(3n+1)}  + \v_{1,0} h_1^{(n)} + \v_{2,0} h_2^{(n)} ,
\\
\bb^{(3n+2)} &= E^{(3n+2)}
- \left(\v_{1,0} +\v_{2,0} \right)E_{\alpha_2}^{(n)}
+ \left(\v_{1,0}-2 \v_{2,0} \right)E_{-(\alpha_1+\alpha_2)}^{(n+1)} \nonu \\
&  +  \frac{1}{3} \left(\v_{1,0} +\v_{2,0} \right) 
\left(\v_{1,0}-2 \v_{2,0} \right)  h_1^{(n)}
+ \frac{2}{3}  \left(\v_{1,0} +\v_{2,0} \right) 
\left(\v_{1,0}-2 \v_{2,0} \right)  h_2^{(n)} .
\label{bcompli}
\end{align}
\end{subequations}
Similarly to  \eqref{commEE}, these
elements form therefore an abelian subalgebra up to 
a central term:
\be \label{abelBB}
\[ \bb^{(3n+p)} ,  \bb^{(3m+p')} \] = 0
\qquad (p, p' = 1,2) .
\ee
The Lax operators 
$\mathcal{L}_N^{\mkdvII} = \pa_{t_N} + A_{t_{N}}^{\mkdvII}$   
admitting nonzero vacuum are indexed 
by  positive flows $N = 3n+ p$, for $p=1,2$ and $n=0,1,2,\dotsc$,
and by negative flows $N = -3n$, for $n=1,2,\dotsc$ --- see 
sec.~\ref{sec:vacslr} below where we discuss this in more generality 
for $\widehat{s\ell}(r+1)$.
Each of these Lax operators obey a zero curvature equation with $\mathcal{L}_1^{\mkdv}$ by definition (this is how the individual
differential equations were constructed to begin with).
Therefore, setting the fields at the vacuum, 
$\v_{i} \to \v_{i,0}$,
these Lax operators obey zero curvature equations
with the operator $\mathcal{L}_{1,\vac}^{\mkdvII}$ in eq.~\eqref{BB}, i.e., $\big[ \bb, A_{t_{N,\vac}}^{\mkdvII}  \big] = 0$.
We conclude that $A_{t_N,\vac}^{\mkdvII} \in \mathcal{K}_\bb$ must 
be a linear combination of the operators in the
abelian subalgebra \eqref{abelBB} and hence commute among themselves.
\footnote{\label{foot_lax_compli}For instance, some of the Lax operators are explicitly given by
\begin{equation*}
\begin{split}
\mathcal{L}_{2,\vac}^{\mkdvII} &= \pa_{t_{2}}+\v_{0,2} \bb^{(1)}+ \bb^{(2)}, \\
\mathcal{L}_{4,\vac}^{\mkdvII} &= \pa_{t_{4}} -\tfrac{1}{3}  \v_{0,2}  \left(-2 \v_{0,1}  \v_{0,2}  +2  \v_{0,1}  ^2+  \v_{0,2}  ^2\right)\bb^{(1)}+\tfrac{1}{3} \left( \v_{0,1} \v_{0,2} - \v_{0,1}^2- \v_{0,2} ^2\right) \bb^{(2)}+ \bb^{(4)} , \\
\mathcal{L}_{-3,\vac}^{\mkdvII} &= \pa_{t_{-3}} +\tfrac{1}{2 \v_{0,1} - \v_{0,2} }\bb^{(-2)}+\tfrac{3 \left( \v_{0,1}  - \v_{0,2}  \right)}{\left( \v_{0,1}  -2  \v_{0,2}  \right) \left(2  \v_{0,1} -  \v_{0,2}  \right) \left( \v_{0,1}  + \v_{0,2} \right)} \bb^{(-1)} .
\end{split}
\end{equation*}
They do not have a simple form and it is unclear whether they follow a well-defined  pattern,
however  each term has the same degree if we associated degree 1 to $\v_{0,i}$ --- recall the discussion
after eq.~\eqref{Bdef} --- i.e., powers of the vacuum times the index of $\bb^{(\cdot)}$ match the
index of the time flow $t_N$.
}
This shows that
$\big[  \mathcal{L}_{N,\vac}^{\mkdvII} , \mathcal{L}_{M,\vac}^{\mkdvII} \big] = 0$
for such flows, which is also true for a general field configuration
in the orbit of this nonzero vacuum by the dressing action.
Moreover, the gauge-Miura correspondence \eqref{gaugesl3} then implies that 
the associated KdV flows in the orbit of a nonzero vacuum 
also commute. In this case the mapping is illustrated as
($n=0,1,\dotsc$,  $m=1,2,\dotsc$)
\be
\begin{tikzcd}[row sep=0.2em, column sep=2em]
t_{3n+1}^{\mkdvII} \arrow[r, "S"] & t_{3n+1}^{\kdv} \\
t_{3n+2}^{\mkdvII} \arrow[r, "S"] & t_{3n+2}^{\kdv} 
\end{tikzcd} \qquad \text{and}
\qquad
\begin{tikzcd}[row sep=0.2em, column sep=2em]
t_{-3m}^{\mkdvII} \arrow[r, "S"] & t_{ -3m+2 }^{\kdv} 
\end{tikzcd} 
\ee

\section{Extension to $\widehat{s\ell}(r+1)$}
\label{sec:extension2}

In light of the previous results we now lay out the extension 
for  general affine Lie algebras $\widehat{s\ell}(r+1)$.
In this case 
the spatial gauge potential of the \emph{multicomponent mKdV 
hierarchy} is given by
\be
A_{x}^{\mkdv} = E + A_0,  \label{mkdvAxslr}
\ee
where 
\be
A_0 = \v_1 h^{(0)}_1 + \dotsm + \v_{r} h^{(0)}_r \in \lie_{0}
\ee
contains $r$ fields $\v_i(x,t_{N})$ ($i=1,\dotsc,r$).
Similarly, the spatial gauge potential of the \emph{multicomponent KdV
hierarchy} is
\be
A_{x}^{\kdv} = E + A_{-1} + \dotsm + A_{-r} \label{kdvAxslr} ,
\ee
where 
\be
A_{-i} = \u_i \; E_{-(\a_1 + \a_2 + \dotsm + \a_i)}^{(0)} 
\in \lie_{-i}
\ee
contains the field $\u_i(x,t_{N})$ ($i=1,\dotsc,r$).
Such hierarchies have positive and negative flow evolution equations as described below.

\subsection{Positive flows}
The positive part of the $\widehat{s\ell}(r+1)$-mKdV hierarchy has
temporal gauge potential in the form 
\be\label{atmkdvmulti}
A_{t_{N}}^{\mkdv} = D^{(N)}_{N} + D^{(N-1)}_{N} + \dotsm + D^{(0)}_{N} ,
\ee
while for the positive part of the $\widehat{s\ell}(r+1)$-KdV hierarchy
we have the form
\be\label{atkdvmulti}
A_{t_{N}}^{\kdv} = \mathcal{D}^{(N)}_{N} + \mathcal{D}^{(N-1)}_{N} + \dotsm + \mathcal{D}^{(-r + 1)}_{N} +  \mathcal{D}^{(-r)}_{N}.
\ee
Thus, the highest grade component of the zero curvature condition
yields 
\be\label{multi_highest}
\[E, D^{(N)}_N\] = 0, \qquad \[E, \mathcal{D}^{(N)}_N\] = 0 ,
\ee
$\pa_x D^{(N)}_N = 0$, and $\pa_x \mathcal{D}^{(N)}_N = 0$,
implying that $N = (r+1)n + p $ for $n=0,1,\dotsc$ and $p = 1,\dotsc,r$ for both hierarchies.
This condition arises because $D^{(N)}_N$ and $\mathcal{D}^{(N)}_N$ must be in the kernel
$\mathcal{K}_E$ defined in eq.~\eqref{kernel.E1.geral}.

\subsection{Negative flows}
The negative  flows of the $\widehat{s\ell}(r+1)$-mKdV follows the structure
\be\label{atmkdvmultineg}
A_{t_{-N}}^{\mkdv} = D^{(-N)}_{-N} + D^{(-N+1)}_{-N} + \dotsm + D^{(-1)}_{-N},
\ee
while for its KdV counterpart  we  propose
\be\label{atkdvmultineg}
A_{t_{-N}}^{\kdv} = \mathcal{D}^{(-N - 2r)}_{-N} + \mathcal{D}^{(-N - 2r + 1)}_{-N} + \dotsm + \mathcal{D}^{(-1)}_{-N} .
\ee
From the zero curvature condition, the lowest grade component 
with gauge potential 
\eqref{atmkdvmultineg} implies no restriction on $N$, i.e., 
$D^{(-N)}_{-N}$ does not need to be in the kernel $\mathcal{K}_E$
and therefore can assume
all negative integer values
$-N = -1, -2,\dotsc$.
However, for the KdV case with potential \eqref{atkdvmultineg} we have the condition
\be
\[ A_{-r}, \mathcal{D}^{(-N - 2r)}_{-N}  \] = 0 ,
\ee
%
which 
together with consistency of the zero curvature condition
requires that $-N = (r+1) n + r$ for $n=-1,-2,\dotsc$.
This is because $\mathcal{D}^{(-N-2r)}_{-N} $  must proportional to the element
$E_{-(\alpha_1 + \dotsm + \alpha_r)}^{(n+1)} \in \lie_{(r+1)n + 1}$ 
so that the zero curvature
condition can be solved non trivially --- here  we again adopt the convention that the first 
negative flow is labelled as $t_{-1}$, 
which requires shifting $-N = (r+1)n + r$.
Therefore, a ``block'' of $r+1$ models of the negative $\widehat{s\ell}(r+1)$-mKdV hierarchy
maps into a single model of the $\widehat{s\ell}(r+1)$-KdV hierarchy via gauge-Miura transformations,
as described in more detail below.
We note also that the $t_{-1}$ mKdV flow corresponds to an affine
Toda field theory which is thus related to the $t_{-1}$ KdV flow by
gauge transformations.

\subsection{Gauge-Miura maps} It has been shown that for $\widehat{s\ell}(r+1)$ the
number of possible gauge-Miura transformations, $S_j$, is given
by $r+1$ \cite{Lobo_2021}. More specifically:
\begin{itemize}
\item We have the form  $S_{j+1}=E^{(-j)}+s^{(-j-1)}+s^{(-j-2)}+\cdots+s^{(-j-r)}$  with $j = 0,1, \ldots, r$, where $E^{(-j)} \in \mathcal{K}_E \in \lie_{-j}$ and $E^{(0)} = I$.
The ``$s$-terms'' can be determined by explicitly solving
the gauge transformations $A_\mu^{\kdv} = S A_\mu^{\mkdv} S^{-1} + S \pa_x S^{-1}$ ($\mu=x,t_{\pm N}$).
\item Each $S_j$ generates one possible Miura-type transformation that connects 
the mKdV fields
$\v_i$ to the KdV fields $\u_i$ ($i=1,\dotsc,r$).
\end{itemize}
Thus, each  gauge transformation $S_j$ maps
one mKdV model into its KdV counterpart. 
For the \emph{positive flows} this correspondence is 1-to-1.
However, for the \emph{negative flows}
this correspondence  is $(r+1)$-to-1, i.e.,
$r+1$ negative mKdV models coalesce  into a single
negative KdV model. This generalizes
the correspondences \eqref{corresp_neg} and \eqref{corresp3}
and is visualized as
\be\label{corresp4}
\begin{tikzcd}[row sep=0.4em, column sep=3em]
t_{N}^{\mkdv} \arrow[r, "S"] & t_{N}^{\kdv} \\
\\
t_{-N}^{\mkdv} \arrow[dr, "S"] & \\
~~~\vdots~~~ \arrow[r, "S"]  &  t_{-N}^{\kdv} \\
t_{-N-r}^{\mkdv} \arrow[ur, swap, "S"] & \\
\end{tikzcd}
\ee
where $S \in \{S_1, \, \dotsc, \, S_{r+1}\}$.
Recall that each of these operators induce one
possible Miura transformation among the fields. Therefore, 
 each mKdV solution generates $r+1$ different
KdV solutions, i.e.,
there exists a large
degeneracy and a rich map regarding solutions of these
hierarchies.%
\footnote{The situation is similar to the plots in figs.~\ref{plot1}
and \ref{plot2} where we obtained a peakon and a 
dark soliton for a KdV model starting from a single solution of an associated mKdV
model.}  
This degeneracy  
is  amplified for the negative 
flows 
since $r+1$ different mKdV models map into a single
 KdV model, i.e.,  now there exists an
$(r+1)\times (r+1)$-fold degeneracy on a solution level.

\subsection{Zero and nonzero vacuum} \label{sec:vacslr}
For the positive flows of the generalized mKdV hierarchy the highest grade component
$\big[E^{(1)}, D^{(N)}_{N,\text{vac}}\big] = 0$
imposes no restriction on $N$, i.e., it matches the time flow indices.
Thus,  all positive models admit both zero and
nonzero vacuum.
For the negative flows, however, the lowest grade component with
a zero vacuum, $\v_i \to \v_{i,0} = 0$ ($i=1,\dotsc,r$), 
is $\big[ E^{(1)}, D_{-N,\text{vac}}^{(-N)}\big] = 0$.
This implies that $D^{(-N)}_{-N} \in \mathcal{K}_E$ so 
the only negative flows 
that admit zero vacuum  are the ones
associated to $-N = (r+1)n + p$, where
$p = 1,\dotsc,r$ is the exponent of the algebra and $n=-1,-2,\dotsc$.
In addition,
in the presence of a nonzero vacuum, $\v_{i} \to \v_{0, i} \ne 0$  ($i=1,\dotsc,r$), the lowest grade component is  given by
$\big[ A_0, D^{(-N)}_{-N}\big] = 0$, implying that  
$D^{(-N)}_{-N} \in \lie_{(r+1)n}$, i.e., $-N = (r+1)n$ for $n=-1,-2,\dotsc$.
These are the models of the negative part of the generalized mKdV hierarchy
that admit nonzero vacuum configuration.
For reference, we summarize these facts:
\begin{itemize}
\item All positive  flows  of the $\widehat{s\ell(}r+1)$-mKdV 
hierarchy admit simultaneously  zero and nonzero vacuum solutions.
\item For an $(r+1)$-block of  negative flows of the $\widehat{s\ell}(r+1)$-mKdV hierarchy the
only model that admits nonzero vacuum solution is the one with index 
$-N=(r+1) n$ ($n=-1,-2,\dotsc$).
The remaining $r$ models associated to $-N = (r+1)n + p$, with $p=1,\dotsc,r$, all have  zero vacuum solutions only.
\end{itemize}

\subsection{Commuting flows}

From the discussion in sec.~\ref{sec:heis}
and sec.~\ref{sec:commflow_sl3}  
it is clear that 
the flows of the generalized mKdV hierarchy commute as a consequence of
the existence of an abelian subalgebra (up to a central term).
In the  \emph{zero vacuum} configuration $\v_i \to \v_{i,0} = 0$ (mKdV-I)
this abelian subalgebra is determined by the kernel  of
the semisimple element $E$ in eq.~\eqref{mkdvAxslr}.
Such elements are explicitly 
given by eqs.~\eqref{kernel.E1.geral} and
\eqref{kerE1elem} and obey
\be \label{abelslr}
\[ E^{((r+1)n + p)}, E^{((r+1)m + p')}\] = 0,
\ee
where  $n,m=0,\pm 1, \dotsc$ and $p,p'=1,\dotsc,r$.

In a \emph{nonzero vacuum} configuration
$\v_i \to \v_{i,0} \ne 0$ ($i=1,\dotsc,r)$
the spatial Lax operator at the vacuum has
the form $\mathcal{L}_{x,\vac}^{\mkdvII} = \pa_x + \bb$,
where
$\bb \equiv E + \v_{1,0} h_1^{(0)} + \dotsm + \v_{r, 0} h_r^{(0)}$.
Now $\bb$ plays the role of the semisimple element $E$ and
its  kernel is a set of elements
$\mathcal{K}_\bb \equiv \{ \bb^{((r+1)n + p)} \}$ obeying
\be
\[ \bb^{((r+1)n + p)}, \bb^{((r+1)m + p')} \] = 0 .
\ee
This  generalizes the elements \eqref{kerE1elem} --- recovered when $\v_{i,0} \to 0$ --- and the commutation relations \eqref{abelslr} to incorporate  a nonzero
vacuum, in close analogy to \eqref{kernelB}--\eqref{abelBB}.
However,  such elements do not seem to have a simple form  (note that  \eqref{bcompli}
already has a complicated form, and even more so for the elements in footnote~\ref{foot_lax_compli}).
Nevertheless,  by the same argument used below eq.~\eqref{abelBB},
the relevant Lax operators at the vacuum must be expressed
as a linear combination of such elements, which then implies
commutation of the general flows of the mKdV hiearchy in the
orbit of this nonzero vacuum (mKdV-II).
The associated flows of the KdV hierarchy also commute
as a consequence of gauge-Miura   \eqref{gaugesl3}.

\section{Conclusions}
\label{sec:conclusion}

We  considered the correspondence between the generalized --- or multicomponent --- mKdV 
hierarchy obtained
from a zero curvature formalism  with the affine Lie algebra 
\mbox{$\widehat{s\ell}(r+1)$} and
the generalized KdV hierarchy following   similar construction. 
There exists $r+1$  gauge-Miura transformations connecting them.
While the positive flows of these hierarchies
(i.e., nonlinear integrable models) are gauge-related in
a 1-to-1 fashion, 
the \emph{negative flows} are related in a \emph{degenerate} $(r+1)$-to-1 fashion, namely $r+1$ models of the generalized mKdV hierarchy are mapped  into a single model of the 
generalized KdV  hierarchy; see  diagram~\eqref{corresp3}.
Moreover, this is true for all  $r+1$ possible 
gauge transformations,
each inducing one Miura-type  transformation among
 the fields. Thus, on a solution level, 
 there is an $(r+1)$-fold degeneracy for
the models within the positive part of these hierarchies, and an $(r+1)\times (r+1)$-fold
degeneracy for the models within the negative part. Thus, 
a single mKdV solution   
generates multiple KdV solutions. 
These results apply to the standard mKdV and KdV hierachies
as a particular case $(r=1)$.

Relationships between  integrable models is of great interest.
Given the importance of the original Miura tranformation in the development of the
inverse scattering transform, which obviously has many ramifications into quantum integrability and 2D CFTs,
the results presented in this paper 
provide a significant generalization thereof besides
uncovering a rich structure for the negative part of  integrable
hierarchies.  We expect that  these  connections may
play a fundamental role in the theory of integrable systems.

The gauge-Miura transformations also give rise to additional
equations for the negative flows, as illustrated explicitly 
 for $\widehat{s\ell}(2)$  as well as for
$\widehat{s\ell}(3)$.
These type of ``temporal Miura transformations'' are the reason
why  $(r+1)$ negative mKdV models are mapped into a single negative KdV model.
These relations are not present for the positive part of these 
hierarchies because there is no degeneracy in the gauge transformation.

The first negative flow of the $\widehat{s\ell}(r+1)$-mKdV hierarchy
corresponds to a relativistic affine Toda field theory 
\cite{Olive_1993,Olive_1993b,Braden_1990,Mikhailov_1981}. 
For instance, for $\widehat{s\ell}(2)$
it is the sinh-Gordon model \eqref{sg_eq}, and  for $\widehat{s\ell}(3)$
it is the model~\eqref{mkdvsl3_minus1}, which can be reduced to the
Tzitz\' eica-Bullough–Dodd model (see sec.~\ref{sec:BD}).
 As we have shown, there are ``KdV counterparts'' to such models connected by gauge tranformations; for  $\widehat{s\ell}(2)$ it is the model~\eqref{kdv_neg1} and for $\widehat{s\ell}(3)$ it is the model~\eqref{kdvsl3_minus1}.
Toda field theories can be defined  however for any affine  Lie 
algebra, and they can be seen as an integrable perturbation of a CFT.
An obvious question thus concerns how such a gauge-Miura correspondence
would play out for algebras beyond $\widehat{s\ell}(r+1)$.
The KdV version of the Toda field theory  would be connected not only
to the latter but to other lower negative flows
of its integrable hierarchy, inheriting solutions from all of these models.
Thus, the ``KdV-Toda model'' can  have many different types
of solution and be able to describe rich nonlinear phenomena.

Another interesting feature of different negative mKdV models is that they
separately admit solutions with a zero  or nonzero vacuum;
they are constituents  of separate integrable hierarchies of commuting
flows defined in the orbit of the vacuum, which  we distinguished
as mKdV-I and mKdV-II, respectively (see sec.~\ref{sec:heis}).
Hence, 
\emph{different} negative mKdV models yield
qualitatively different
types of solutions to the \emph{same} negative KdV 
model which they are related to. 
This was illustrated explicitly for  $\widehat{s\ell}(2)$ 
and proved abstractly for $\widehat{s\ell}(r+1)$.
This is in contrast to the positive part of these hierarchies where
each model admits simultaneously zero and nonzero vacuum solutions and
are 1-to-1 related.
We illustrated soliton solutions, or more precisely \emph{peakons} and
\emph{dark-solitons}. However, the gauge-Miura
transformations are general and one can also consider 
quasi-periodic --- or finite-gap --- solutions.
Such solutions can be expressed in terms of theta-functions
\cite{Krichever_1980,Dubrovin_1981,Date_1982} and have deep connections 
in algebraic-geometry and Riemann surfaces. 
Quasi-periodic
solutions are known for several standard integrable models, such
as KdV, mKdV, and sinh-Gordon. 
It would  be  interesting to consider quasi-periodic solutions of more
complicated 
 models such as \eqref{mkdvt2sl3}--\eqref{mkdvsl3_minus3}, and potentially
 other models from $\widehat{s\ell}(r+1)$, 
 besides understanding
 how they arise from a suitable vacuum; perhaps they
 generate a ``Type-III'' hierarchy of  commuting flows in 
 the orbit of such a vacuum 
 according to our perspective.
 (We plan to present  connections specific to quasi-periodic solutions elsewhere.)

\bigskip

\subsubsection*{Acknowledgements}
\vspace{-1em}
We are indebted to the referee from JHEP for
posing the important question about  commuting flows
of integrable hierarchies, besides careful analysis
of the paper. 
JFG and AHZ thank CNPq and FAPESP for support. YFA thanks FAPESP for financial support under grant \#2021/00623-4 and \#2022/13584-0. GVL is supported by CAPES. GF thanks UC Berkeley, where this work was partially completed, and 
in particular MI Jordan for support. 
This research was financed in part by 
CAPES (finance Code 001).

\appendix

\section{Affine Lie algebras} 
\label{sec:algebra}

\subsection{$\widehat{s\ell}(r+1)$}
\label{sec:slr}

We follow the conventions and notation
from \cite{Olive_1985b,Cornwell_1989} and state
only the relevant relations for our purposes.
Consider the affine Lie algebra $\lie \equiv \widehat{s\ell}(r+1)$, which is the 
the Kac-Moody algebra  $A_r$ without
the central extension. The  generators of the algebra are
denoted by
\begin{equation}
\begin{split}
\lie = \Big\{ 
& h_1^{(m)},  \dotsc,  h_r^{(m)}, 
 E_{\pm\a_1}^{(m)},  \dotsc,  E_{\pm\a_r}^{(m)},  
 E_{\pm(\a_1+\a_2)}^{(m)}, \dotsc, 
E_{\pm(\a_{r-1}+\a_r)}^{(m)}, \\ 
& \dotsm ,  \,
 E_{\pm(\a_1+ \dotsm + \a_{r-1})}^{(m)}, 
 E_{\pm(\a_2+ \cdots + \a_{r})}^{(m)}, \; 
E_{\pm(\a_1+ \cdots + \a_{r})}^{(m)} \Big\} ,
\end{split}
\end{equation}
where 
$\a_j$ is a  
simple root. 
We also have
$\big(E^{(m)}_{\alpha_i}\big)^\dagger = E_{-\alpha_i}^{(-m)}$
and $\big(h^{(m)}_i\big)^\dagger = h^{(-m)}_i$.
The commutation table reads
\begin{subequations}
\begin{align}
\left[h_i^{(m)},h_j^{(n)}\right] &=   0 , \\
\left[h_i^{(m)}, E_{\alpha_j}^{(n)}\right] &= (\alpha_i \cdot \alpha_j)   E_{\alpha_j}^{(m+n)} , \\
\left[E_{\alpha}^{(m)}, E_{\beta}^{(n)}\right] &=  
\begin{cases} 
\epsilon(\alpha, \beta)   E_{\alpha + \beta}^{(m+n)} &  \mbox{if $\alpha +\beta$ is a root,} \\
 \alpha \cdot H^{(m+n)}  & \mbox{if $\alpha + \beta=0$,} \\
0 & \mbox{otherwise. }
\end{cases}
\end{align}
\end{subequations}
The \emph{principal grading operator} is  defined by 
\be
\widehat{Q} \equiv (r+1) \widehat{d} + 
\left( \mu_1 + \cdots + \mu_r \right) \cdot H
\label{Q} ,
\ee
where $h^{(m)}_i = \a_i \cdot H^{(m)}$ and
 $\widehat{d}$ is the derivation operator defined as
\be
\left[ \widehat{d},  T^{(m)} \right] = m \; T^{(m)}
\ee
for $T^{(m)} \in \lie$,
and $\mu_i$ are the \emph{fundamental weights} obeying
\be
\left[ \mu_i \cdot H^{(m)} ,   E_{\a_j}^{(n)} \right] = 
\left( \mu_i \cdot \a_j  \right) E_{\a_j}^{(m+n)} 
\ee
with $\mu_i \cdot \a_j = \delta_{i,j}$ for $i,j = 1,2, \dots, r$. 
This operator decomposes the  algebra  into graded subspaces,
$\mathcal{G}  
= \sum_a \lie_a$, where
\begin{equation}
\left[\widehat{Q}_, \mathcal{G}_a\right]= a  \mathcal{G}_a, 
\qquad
\left[\mathcal{G}_a, \mathcal{G}_b\right] \subset \mathcal{G}_{a+b}, 
\label{P}
\end{equation}
for $a, b \in \mathbb{Z}$. 
From \eqref{Q} and \eqref{P} it follows that
\begin{subequations}
\label{graded_subspaces}
\begin{align}
\lie_{(r+1)m} &= \left\{ h_1^{(m)}, \dots, h_r^{(m)}   \right\}, \\
\lie_{(r+1)m+1} &= \left\{ E_{\a_1}^{(m)}, \dots, E_{\a_r}^{(m)}, E_{-(\a_1 + \cdots + \a_r)}^{(m+1)}   \right\}, \\
\lie_{(r+1)m+2} &= \left\{ E_{\a_1+\a_2}^{(m)}, \; E_{\a_2+\a_3}^{(m)},  \dots, E_{\a_{r-1}+\a_{r}}^{(m)}, \; E_{-(\a_1 + \cdots + \a_{r-1})}^{(m+1)}, \; E_{-(\a_2 + \cdots + \a_r)}^{(m+1)}   \right\}, \\
&~~\vdots  \nonu \\  
\lie_{(r+1)m+r} &= \left\{ E_{-\a_1}^{(m+1)}, \dots, E_{-\a_r}^{(m+1)}, E_{\a_1 + \cdots + \a_r}^{(m)}   \right\}.
\end{align} 
\end{subequations}
The semisimple element  
\begin{equation} \label{Edef}
        E \equiv E_{\a_1}^{(0)} + E_{\a_2}^{(0)} + \cdots + E_{\a_r}^{(0)} + E_{-(\a_1 + \cdots + \a_r)}^{(1)}
\end{equation}
provides a  kernel-image decomposition,
$ \lie  = {\cal K}_{E} \oplus {\cal M}_{E}  $, where 
$\mathcal{K}_E = \{  X \in {\cal {K}}_E \, | \, [X, E]=0 \}$ and 
$\mathcal{M}_E = \lie \setminus \mathcal{K}_E$. Due to the form of \eqref{Edef} and \eqref{graded_subspaces}
one finds
\begin{equation}
\mathcal{K}_E = \left\{ E^{(r+1)m+1}, E^{(r+1)m+2}, \dots, E^{(r+1)m+r} \right\} ,
\label{kernel.E1.geral}
\end{equation}
where
\begin{subequations} \label{kerE1elem}
\begin{align}
E^{(r+1)m+1} &= E_{\a_1}^{(m)} + E_{\a_2}^{(m)} + \cdots + E_{-(\a_1 + \cdots + \a_r)}^{(m+1)},          \\
E^{(r+1)m+2} &= E_{\a_1+\a_2}^{(m)} + E_{\a_2+\a_3}^{(m)} + \cdots + E_{\a_{r-1}+\a_{r}}^{(m)} + E_{-(\a_1 + \cdots + \a_{r-1})}^{(m+1)} + E_{-(\a_2 + \cdots + \a_r)}^{(m+1)},         \\
&~~\vdots  \nonu \\
E^{(r+1)m+r} &= E_{-\a_1}^{(m+1)} + E_{-\a_2}^{(m+1)} +  \dots + E_{-\a_r}^{(m+1)} + E_{\a_1 + \cdots + \a_r}^{(m)}.
\end{align}
\end{subequations}

\subsection{$\widehat{s\ell}(2)$}
\label{sec:sl2}

As a particular case of the previous construction we have
the affine Lie algebra $\lie \equiv \widehat{s\ell}(2)$
with commutation relations
\begin{equation} \label{comm_sl2}
\begin{split}
\left[h^{(m)}, h^{(n)} \right] = 0, \qquad
\left[h^{(m)}, E_{\pm \alpha}^{(n)}\right] = \pm 2 E_{\pm \alpha}^{(m+n)}, \qquad
\left[E_\alpha^{(m)}, E_{-\alpha}^{(n)}\right] = h^{(m+n)} ,
\end{split}
\end{equation}
and principal grading operator 
$ \widehat{Q} \equiv 2 \widehat{d}  +\frac{1}{2} h^{(0)}$
yielding  
\be 
\mathcal{G}_{2 n} = \big\{h^{(n)}\big\},  \qquad
\mathcal{G}_{2 n+1} =\big\{E_\alpha^{(n)} + E_{-\alpha}^{(n+1)}\big\}. 
\label{graded_subspaces2}
\ee
The semisimple element  is chosen as
\begin{equation} \label{Edef2}
 E \equiv E_{\alpha}^{(0)} + E_{-\alpha}^{(1)} \in \lie_1 ,
\end{equation}
which defines the kernel subspace
\begin{equation} \label{Kernel}
    \mathcal{K}_E = 
    \big\{ E_\alpha ^{(n)} + E_{-\alpha}^{(n+1)} \big\}  \in \lie_{2n+1} . 
\end{equation}

\subsection{$\widehat{s\ell}(3)$}
\label{sec:sl3}

For 
$\lie \equiv \widehat{s\ell}(3)$ we have  
the commutation table
\begin{subequations}
\begin{align}
\left[ h_i^{(m)}, E_{\a_j} ^{(n)} \right] &=\a_i \cdot \a_j \, E_{\a_j} ^{(m+n)}, \\  
\left[ h_i^{(m)},h_j^{(n)} \right] &=0, \\
\left[E_{\a_i}^{(m)},E_{-\a_i}^{(n)}\right] &= h_{i}^{(m+n)}, \\    
\left[ E_{\a_1}^{(m)}, E_{\a_2}^{(n)} \right] &= E_{\a_1+\a_2}^{(m+n)} .
\end{align}
\end{subequations}
The principal grading  operator 
$ 
\widehat{Q} = 3 \widehat{d}  +h_1^{(0)} + h_2^{(0)} $
yields
\begin{subequations} \label{sl3grade}
\begin{align}
\lie_{3m} &= \big\{h_1^{(m)},\, h_2^{(m)} \big\} , \\
\lie_{3m+1} &= \big\{ E_{\alpha_1}^{(m)},\, E_{\alpha_2}^{(m)}, \, E_{-(\alpha_1+\alpha_2)}^{(m+1)}\big\}, \\
\lie_{3m+2} &=   \big\{   E_{-\alpha_1}^{(m+1)},\, E_{-\alpha_2}^{(m+1)},\,E_{(\alpha_1+\alpha_2)}^{(m)}  \big\}.
\end{align}
\end{subequations}
Finally, the semisimple element 
\begin{equation} 
E \equiv E_{\alpha_1}^{(0)}
+E_{\alpha_2}^{(0)} + E_{-(\alpha_1+\alpha_2)}^{(1)} 
\end{equation}
defines the kernel
\begin{equation} \label{Kernelsl3}
    \mathcal{K}_E = \{
    E^{(3m+1)}, E^{(3m+2)} \} ,
\end{equation}
where
\be
\begin{split}
E^{(3m+1)}&= E_{\alpha_1}^{(m)}+E_{\alpha_2}^{(m)}+E_{-(\alpha_1+\alpha_2)}^{(m+1)} \in \lie_{3m+1},\\
E^{(3m+2)}&= E_{-\alpha_1}^{(m+1)}+E_{-\alpha_2}^{(m+1)}+E_{(\alpha_1+\alpha_2)}^{(m)} \in \lie_{3m+2}.
\end{split}
\ee

\section{Alternative gauge transformation}
\label{sec:alternative_gauge}
In sec.~\ref{sec:gauge_neg_flows} we considered the gauge transformation with
the operator $S_1$ --- see eq.~\eqref{g} --- which induces a Miura transformation \eqref{miura2}
with \emph{minus sign}. Alternatively, one can consider $S_2$ which
induces a Miura transformation with \emph{plus sign} instead (see also \cite{Lobo_2021}
for details).
For completeness, we discuss this case in this section.

The matrix representation of $S_2$ was given in eq.~\eqref{miura_gauge}, which 
we repeat for convenience:
\be \label{m2}
S_2 = \begin{pmatrix} 0 & \lambda^{-1} \\ 1 & - \lambda^{-1} \pa_x \phi \end{pmatrix}, 
\qquad 
\pa_x\eta = (\pa_x \phi)^2 + \pa_x^2 \phi .
\ee
By comparison with the definition of $S_1$ one concludes that
\begin{equation}
S_2(\phi)= S_1(-\phi) E^{(-1)} , \qquad E^{(-1)} \equiv E_{\a}^{(-1)} + E_{-\a}^{(0)} .
\end{equation}
A gauge transformation using $S_2$ is written as ($\mu=x,t_N$)
\be\begin{split}
A_{\mu}^{\kdv}&= S_2(\phi) A_{\mu}^{\mkdv} S_2^{-1}(\phi)+S_2 (\phi) \partial_{\mu} S_2^{-1} (\phi) \\
&= S_1(-\phi)  E^{(-1)} A_{\mu}^{\mkdv} E^{(1)}  S_1^{-1}(-\phi) +S_1(-\phi) E^{(-1)}\partial_{\mu} (E^{(1)}  S_1^{-1}(-\phi)) \\
&= S_1(-\phi)  E^{(-1)} A_{\mu}^{\mkdv} E^{(1)}  S_1^{-1}(-\phi) +S_1(-\phi)\partial_{\mu}  S_1^{-1}(-\phi).
\end{split}
\ee
The following relations are useful in the next derivations:
\be
E^{(-1)} \; h^{(n)} \; E^{(1)} =  -  h^{(n)}, \quad 
E^{(-1)} \; E_{\a}^{(n)} \; E^{(1)} =  E_{-\a}^{(n+1)}, \quad
E^{(-1)} \; E_{-\a}^{(n+1)} \; E^{(1)} =  E_{\a}^{(n)}.
\ee
Thus, gauging $A_x^{\mkdv}(\phi )  =  E_{\alpha}^{(0)} + E_{-\alpha}^{(1)} + \pa_x \phi  h^{(0)}$ via $S_2$ yields
\be \begin{split}
A_{x}^{\kdv} &= S_1(-\phi)  E^{(-1)} \left(E_{\alpha}^{(0)} + E_{-\alpha}^{(1)} + \pa_x \phi \, h^{(0)}\right) E^{(1)}  S_1^{-1}(-\phi) +S_1(-\phi)\partial_{x}  S_1^{-1}(-\phi) \\
&= S_1(-\phi)\left(E_{\alpha}^{(0)} + E_{-\alpha}^{(1)} - \pa_x \phi \, h^{(0)}\right)  S_1^{-1}(-\phi) +S_1(-\phi)\partial_{x}  S_1^{-1}(-\phi) \\
& = S_1(-\phi)A_x^{\mathrm{mKdV}}(-\phi ) S_1^{-1}(-\phi) +S_1(-\phi)\partial_{x}  S_1^{-1}(-\phi).
\end{split}
\ee
Therefore, the  transformation with $S_2$ is equivalent to the transformation
with $S_1$ but with the mKdV field reflected, $\phi \to -\phi$;
this is why the Miura transformation \eqref{m2} picks up a plus sign. 


For the negative part of the hierarchies we need to carefully consider
the gauge transformation for $A_{t_{-N}}$.
Let $N$ be the index of the KdV flow, while 
$M = 2n - 1$ or $M = 2n$ denotes the corresponding odd/even index of the mKdV flow --- 
we use the same notation of eq.~\eqref{phiba}. 
We thus have
\be\begin{split}
A_{\t_{-N}}^{\kdv} &=  S_2(\phi) A_{t_{-M}}^{\mkdv} S_2^{-1}(\phi)+S_2 (\phi) \partial_{t_{-M}} S_2^{-1} (\phi) \\
&= S_1(-\phi)  E^{(-1)} A_{\t_{-M}}^{\mkdv} E^{(1)}  S_1^{-1}(-\phi) +S_1(-\phi)\partial_{t_{-M}}  S_1^{-1}(-\phi).
\end{split}
\ee
Taking the grade $-1$ component on both sides yields 
\be     \begin{split}
\mathcal{D}^{(-1)}_{-M}  &= 
E^{(-1)} D^{(-1)}_{-M} E^{(1)} + \pa_{t_{-M}}\pa_x\phi \, E_{-\a}^{(0)} \\    
& = E^{(-1)} \left(a^{(-1)}_{-M} E_{\a}^{(-1)} +b^{(-1)}_{-M}  E_{-\a}^{(0)}  \right) E^{(1)}+ \pa_{t_{-M}}\pa_x \, E_{-\a}^{(0)} \\
& = b^{(-1)}_{-M}  E_{\a}^{(-1)} + a^{(-1)}_{-M}  E_{-\a}^{(0)} + 
\pa_{t_{-M}} \pa_x \phi  \, E_{-\a}^{(0)} .
\end{split}
\ee
Similarly to the case of eq.~\eqref{etaaminus1} we now conclude instead that
\be
\pa_{t_{-N}} \eta = 2  b^{(-1)}_{-M} [\phi(x, t_{-M})].
\end{equation}
Thus, the effect of $S_2$ compared to $S_1$ is to interchange
$a^{(-1)}_{-M}$ and $b^{(-1)}_{-M}$.
For instance, for the  flow $\t_{-1}$ of the KdV hierarchy, this yields
the same relation \eqref{mt1} but with $\phi \to -\phi$, i.e.,
\be
\pa_{\t_{-1}} \eta = 2 e^{2 \phi(x, t_{-1})},
\ee
and now the Miura transformation with plus sign holds
\eqref{m2}.  Moreover, the relation \eqref{mt2} now becomes
\be \label{m22}
\pa_{t_{-1}} \eta= -4 e^{2\phi(x,t_{-2})} \dI e^{-2\phi(x, t_{-2})} . 
\ee

Such a symmetry, $S_1 \to S_2$ and $\phi \to -\phi$, can be anticipated from the parity invariance of the equations of motion of the mKdV hierarchy. More precisely, 
odd time flows of the mKdV hierarchy are invariant by parity transformation, however the negative even equations pick up a global minus sign. This is why
\eqref{m22} has an additional sign compared to \eqref{m2}, besides reflecting $\phi$.
On the other hand, the equations of motion of the KdV hierachy do not
have such a   symmetry under parity, which is automatically 
corrected by the Miura transformation \eqref{m2}.

\section{Explicit calculation of commuting flows}
\label{sec:examplecomm}

In sec.~\ref{sec:heis} we demonstrated abstractly that the flows within each mKdV-I (zero vacuum)
and mKdV-II (nonzero vacuum) hierarchies commute.
Here we provide an explicit example to illustrate this fact.
We focus on the $t_3$ and ${t_{-2}}$ flows of mKdV-II. The former is the 
mKdV equation \eqref{kdv_mkdv_t3} and the latter is model \eqref{mkdv_minus_two_eq}
that only admits nonzero vacuum solutions.
Denote $\v = \pa_{t_1} \phi(t_{1}, t_{3}, \dotsc; t_{-2}, t_{-4}, \dotsc)$ 
the field of the entire mKdV-II hierarchy, and recall that $t_1 = x$ (chiral wave
equations).
The differential equations of interest are
\be\label{eqt3flow}
4 \pa_{t_3} \v = 4\pa_{t_{3}} \pa_x \phi = \pa_x^4 \phi - 2\pa_x (\pa_x \phi)^3,
\ee
which can also be written as
\be \label{eqt3flow2}
4 \pa_{t_{3}} \phi = \pa_x^3 \phi -2(\pa_x \phi)^3
\ee
by acting with the inverse operator \eqref{dIDef},
and
\be \label{eqtm2flow}
\pa_{t_{-2}} \v = \pa_{t_{-2}}  \pa_x \phi = - 2 \mathcal{A}_+(\phi),
\ee
where, for convenience, we define
\be\label{Aops}
\mathcal{A}_{\pm}(\phi) \equiv  e^{-2 \phi} \dI e^{2 \phi} \pm e^{2 \phi} \dI e^{-2 \phi}  .
\ee

First, note that we can use $\pa_{t_N} \pa_x \phi = \pa_x \pa_{t_N} \phi$ for any
relevant $N$; the differential equations
are constructed from the zero curvature condition
$\big[ \mathcal{L}_x^{\mkdvII}, \mathcal{L}_{t_N}^{\mkdvII}  \big] = 0$, i.e.,
$t_1=x$ commutes with all $t_N$'s  by definition.
Next,  acting with $\pa_{t_{-2}}$ on eq.~\eqref{eqt3flow} yields
\be \label{t2t3}
4\pa_{t_{-2}}\pa_{t_3} \v = 
-2 \partial_x \Big(  \big(\partial_x^2  - 6 (\pa_x \phi)^2 \big) \mathcal{A}_+(\phi) \Big). 
\ee
Note that
\be\label{Aopsder}
\pa_x \mathcal{A}_+ = -2 (\pa_x\phi) A_- + 2, \qquad 
\pa_x \mathcal{A}_- = -2 (\pa_x\phi) A_+ ,
\ee
thus
\begin{subequations}
\begin{align}
\pa_x^3 \mathcal{A}_+ &= -2( \pa_x^3 \phi) \mathcal{A}_-
+12(\pa_x \phi)(\pa_x^2 \phi) \mathcal{A}_+
- 8 (\pa_x \phi)^3 \mathcal{A}_-
+ 8 (\pa_x \phi)^2 , \\
\pa_x\big( (\pa_x \phi)^2 \mathcal{A}_+  \big) &= 2(\pa_x \phi) (\pa_x^2 \phi)\mathcal{A}_+
-2 (\pa_x \phi)^3 \mathcal{A}_- + 2 (\pa_x \phi)^2.
\end{align}
\end{subequations}
Replacing these relations into eq.~\eqref{t2t3} we obtain
\be \label{t2t3_final}
4 \pa_{t_{-2}} \pa_{t_3} \v = 4 \Big( (\pa_x^3 \phi) - 2 (\pa_x \phi)^2  \Big) \mathcal{A}_-(\phi) + 8 (\pa_x \phi)^2 .
\ee

Conversely, hitting with $\pa_{t_3}$ on eq.~\eqref{eqtm2flow} yields
\be
\begin{split}
4 \pa_{t_3} \pa_{t_{-2}} \v &= - 8 \pa_{t_3} \mathcal{A}_+(\phi) \\
& = 
16 (\pa_{t_3} \phi) \mathcal{A}_- (\phi)
-8 e^{-2\phi} \pa_x^{-1} \Big( \pa_{t_3} e^{2\phi}  \Big)
-8 e^{2\phi} \pa_x^{-1} \Big( \pa_{t_3} e^{-2\phi}  \Big).
\end{split}
\ee
Note that upon using eq.~\eqref{eqt3flow2} we have
\be
\begin{split}
2 \pa_{t_3} e^{\pm 2\phi} &=  \pm e^{\pm 2\phi} \Big( \pa_x^3\phi - 2(\pa_x\phi)^3  \Big) \\
&=  \pa_x\Big(  e^{\pm 2 \phi} \big( \pm \pa_x^2 \phi - (\pa_x \phi)^2  \big)  \Big).
\end{split}
\ee
Using this into the previous equation we finally obtain
\be
\begin{split}
4 \pa_{t_3} \pa_{t_{-2}} \v &= 16 (\pa_{t_3} \phi) \mathcal{A}_-(\phi) + 8 (\pa_x\phi)^2  \\
&= 4 \Big( (\pa_x^3 \phi) - 2 (\pa_x \phi)^2  \Big) \mathcal{A}_-(\phi) + 8 (\pa_x \phi)^2 ,
\end{split}
\ee
which is equal to \eqref{t2t3_final}, as desired.

\bibliography{biblio.bib}

\end{document}